\documentclass{emulateapj}
\usepackage{amsmath}
\usepackage{graphicx}
\usepackage{epsfig}

\usepackage{url}
\usepackage{indentfirst}
\begin{document}
\title{High-Energy emissions from Pulsar/Be binary system PSR J2032+4127/MT91~213}
\author{Takata, J.\altaffilmark{1},  Tam, P.H.T.\altaffilmark{2}, Ng, C.W.\altaffilmark{3}, Li., K.L.\altaffilmark{4}, Kong, A.K.H.\altaffilmark{5}, Hui, C,Y\altaffilmark{6}, \and Cheng, K. S.\altaffilmark{3}}
\email{takata@hust.edu.cn}
\altaffiltext{1}{School of physics, Huazhong University of Science and Technology, Wuhan 430074, China}
\altaffiltext{2}{Institute of Astronomy and Space Science, Sun Yat-Sen University, Guangzhou 510275, China}
\altaffiltext{3}{Department of Physics, The University of Hong Kong, Pokfulam Road, Hong Kong}
\altaffiltext{4}{Department of Physics and Astronomy, Michigan State University, East Lansing, MI 48824-2320, USA}
\altaffiltext{5}{Institute of Astronomy and Department of Physics, National Tsing Hua University, Hsinchu, Taiwan}
\altaffiltext{6}{Department of Astronomy and Space Science, Chungnam National University, Daejeon 305-764, Republic of Korea}
\begin{abstract}
PSR~J2032+4127 is a radio-loud gamma-ray-emitting pulsar; it is orbiting 
around a high-mass Be type star with a very long orbital period of $25-50$years,
and is approaching periastron, which will  occur
in late 2017/early 2018. This system comprises with a young pulsar and a Be type star, which
is similar to the so-called  gamma-ray binary PSR~B1259-63/LS2883. 
It is expected  therefore that PSR~J2032+4127 shows 
an enhancement of high-energy emission caused by the
interaction between the pulsar wind and Be wind/disk around periastron.
Ho et al. recently reported a rapid increase in the X-ray flux
from  this system.  In this paper,
we also confirm a rapid increase in  the X-ray flux along the orbit,
while the GeV flux shows no significant change.
We discuss the high-energy emissions  from the shock caused by  
the pulsar wind and stellar wind interaction and examine the properties
of the pulsar wind in this binary system.
 We argue  that the rate of increase   of  the X-ray flux  observed by Swift
indicates (1) a variation of the momentum ratio of the two-wind interaction region along
the orbit, or (2) an evolution of
 the magnetization parameter of the pulsar wind
 with the radial distance from the pulsar. We also discuss
the  pulsar wind/Be disk interaction
at the periastron passage, and  propose the possibility of formation of an  accretion 
disk around the pulsar. We model  high-energy  emissions through the inverse-Compton scattering 
process of the  cold-relativistic pulsar wind off  soft photons from the accretion disk. 
\end{abstract}

\section{Introduction}
\begin{figure}
  \centering
  \epsscale{1.0}
  \plotone{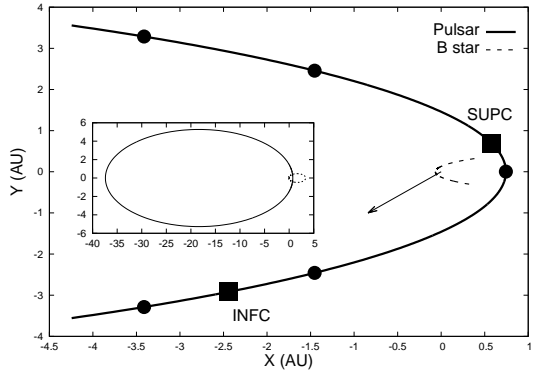}
  \caption{Schematic diagram illustrating the orbit of PSR J2032+4127 (solid)
    and  B star companion MT91~21 (dashed line) around the periastron,
    viewed from perpendicular to the orbit plane.
    The orbit parameters are $P_{ob}=17000$days and $e=0.961$ (Ho et al. 2016). The main panel
    and inserted panel show the orbit during  -250 days and 250days from the periastron and entire orbit, respectively.
    The filled circles mark 100days interval.
    The arrows indicate the direction to the Earth, and
    the boxes show the positions of the inferior conjunction (INFC) and superior
  conjunction (SUPC), respectively.}
  \label{orbit}
\end{figure}

 PSR J2032+4127 is a radio-loud gamma-ray-emitting pulsar discovered by 
the Fermi Large Area Telescope (Fermi-LAT)  and it is a young 
pulsar with a spin  period $P=143$ms, 
spin-down power $L_{sd}\sim 1.7\times 10^{35}{\rm erg~s^{-1}}$ and 
spin-down age $\tau\sim 180$kyr (Abdo et al. 2013). 
This pulsar had been regarded 
as  isolated pulsar (Camilo et al. 2009) because of the lack of 
 apparent  variation  in its rotation  caused by reasonable 
orbital motion.  Lyne et al. (2015) found  the variation of 
the observed pulsar rotation rate,
which is consistent with the Doppler effect of 
the orbital motion around the high-mass Be star MT91 213. Their study suggests 
that the orbital period is very long, $P_{ob}\sim 25$yr, and the orbit 
is extremely elongated with  an eccentricity $e\sim 0.93$. Ho et al. (2016)
refined the orbit parameters as $P_{ob}\sim 50$years and $e\sim 0.96$
(see Figure~\ref{orbit}).
PSR~J2032+4127 is a member of a class of rare-type pulsar that is orbiting around a
high-mass B star. The next periastron passage will occur 
in late 2017 or in early 2018.

The binary nature of PSR~J2032+4127 is similar to that of PSR~B1259-63, which  is
a well known radio pulsar that is orbiting around 
the high-mass B star, LS2883, with  $P_{ob}\sim 3.4$years
and $e\sim 0.83$ (Johnston et al. 1992, 1999, 2005).
PSR B1259-63/LS2883 is  known to be  a source
of non-pulsed  and non-thermal emissions in the X-ray/GeV/TeV bands
(Aharonian et al. 2005; Abdo et al. 2011; Tam et al. 2011, 2015), and it is classified as a
so-called gamma-ray binary, which comprises a compact object (neutron star
 or black hole)  and a high-mass OB star (see Dubus 2013). 
Multi-wavelength observations have confirmed several TeV gamma-ray binaries, 
namely, PSR B1259-63/LS 2883, LS5039 (Aharonian et al. 2006),
 LS~I+61$^{\circ}$~303 (Albert et al. 2006), 
1FGL~J1018.6-5856 (Ackermann et al. 2012), and
 H.E.S.S.~J0632+057 (Hinton et al. 2009).  PSR~B1259-63/LS2883 is 
the only binary system for which the compact object has been confirmed  
to be a young pulsar.    
PSR~J2032+4127/MT91~213 is a candidate for the  next  gamma-ray binary, 
for which the compact object is definitely  a  young pulsar.

\begin{figure}
  \centering
  \epsscale{1.0}
  \plotone{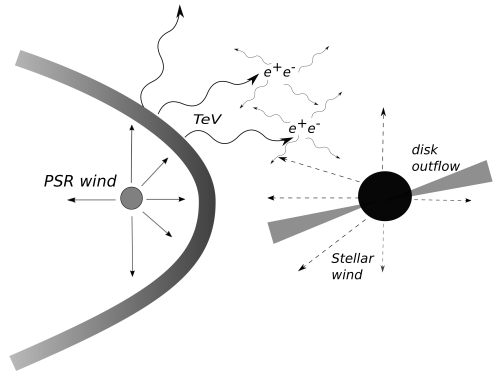}
  \caption{Schematic picture of the interaction of the pulsar wind and Be
    wind/disk. The
    synchrotron process and ICS process of the shocked pulsar wind produce X-rays and TeV gamma-rays, respectively.
    The TeV gamma-rays may create new electron and positron pairs by the photon-photon pair-creation process,
    and they initiate a pair-creation cascade, if they propagate toward the companion star.  }
  \label{system}
\end{figure}

The origin of the high-energy TeV emissions from the PSR~B1259-63/LS~2883
system is likely related to the interaction
of the pulsar wind of PSR~B1259-63
and the Be wind/disk of LS~2883 (Tavani \& Arons 1997; Dubus 2006, 2013). 
Their interaction results in the formation of a shock, where the ram  pressures of the pulsar
wind and of the  stellar wind/disk are in balance (Figure~\ref{system}).
The electrons/positrons in the pulsar wind are accelerated at the shock, and produce nonthermal
emissions in the radio to TeV gamma-ray bands.  The standard scenario assumes the synchrotron radiation
and the inverse-Compton scattering (ICS) process produces radio/X-ray emission
and TeV gamma-ray emission,
respectively.  We  can expect that PSR~J2032+4127/MT91~213 is also a source of   
 X-rays/TeV gamma-rays arising from the   same mechanisms.

 X-ray/TeV emissions from the gamma-ray binaries have been observed to
 exhibit temporal variations in their emission.
 Various models have been suggested  to explain the energy-dependent orbital modulations: for the
 pulsar model of the gamma-ray binaries, for example,
 the Doppler boosting effect due to the finite velocity of
   the shocked pulsar wind (Dubus et al. 2010; Kong et al. 2012; Takata et al 2014a),
   evolution of the energy spectra of relativistic electrons of the shocked pulsar wind
   under the different  energy-loss
   rates (Khangulyan et al. 2007; Takahashi et al. 2009; Zabalza et al. 2013),
   and the  interaction between the pulsar wind and the
   Be disk (Sierpowska-Bartosika \&  Bednarek 2008; Takata et al. 2012), have been suggested. 
   In this paper, we will consider
   the orbital variation of the  emission from  PSR~J2032+4127/MT91~213 as a result of the radial
   evolution of the magnetization parameter (Takata \& Taam 2009; Kong et al. 2011) or the variation of the momentum ratio
   of the two winds.  

  The GeV emission from PSR~J2032+4127/MT91~213 is modulating with the spin period of the pulsar,
  and hence is dominated bythe  magnetospheric process.   The GeV emission from the
  gamma-ray binaries has been observed by the Fermi-LAT.
  For PSR~B1259-63/LS~2883, flare-like GeV emissions have been observed after
  the second Be disk passage
  of the pulsar (Abdo et al., 2011; Tam et al. 2011, 2015).
  Although the inverse-Compton
  scattering process with anisotropic soft-photon field from the disk/companion star  (e.g Khangulyan et al. 2011, 2014; van Soelen et al. 2012;
  Dubus \& Cerutti 2013) or synchrotron radiation
  process of the relativistic electrons/positrons  (Chernyakova et al. 2015; Xing et al. 2016)
  have been considered, the origin of the emission is not yet understood.  

  The gamma-binary LS~5039, whose orbital period is $~3.9$days, is a candidate
  for which the compact object is a young pulsar. The GeV emission from this system has
  been observed  over the whole orbital phase by the Fermi-LAT (Abdo et al. 2009).
  Because their   spectral shape measured by Fermi resembles those of the gamma-ray emitting pulsars,
  it has been suggested that LS 5039 includes a young pulsar.  The orbital
  modulating GeV emission confirmed by Fermi, however, suggests
  the emission process in the intra-binary space. The orbital modulations of GeV and X-ray/TeV emissions from LS~5039 show anti-correlation:
  GeV flux (or X/TeV fluxes)
  becomes maximum around the superior conjunction (hereafter SUPC) (or inferior conjunction (INFC))
  and becomes minimum around the INFC (or SUPC).
  Within the framework of the shock emission due to the interaction
  between the pulsar wind and stellar wind,
  the observed modulation of the X-ray emissions  from gamma-ray
  binary LS~5039 has been interpreted as the Doppler boosting effect caused by the finite velocity of
  the shocked pulsar  wind (Dubus et al. 2010; Takata et al. 2014a),
  while the strong absorption of the TeV photons
  by the stellar photon field  will suppress the observed emissions at around SUPC (Dubus et al. 2008).
  The GeV emission may be as  a result
  of the inverse-Compton scattering
  (hereafter ICS) process of  the cold-relativistic pulsar wind
  (Sierpowska-Bartosik \& Torres 2008; Kapala et al. 2010; Torres 2011; Takata et al. 2014a) or of 
  the shock-accelerated particles (Yamaguchi \& Takahara; Zabalza et al. 2013).
  The ICS process between the relativistic particles
  and the stellar photons causes  a flux peak in the observed
  modulation at around the SUPC (if there is
  no absorption), where the  scattering  process of the relativistic  particles that propagate to the
  Earth occurs the head-on, and causes  a flux minimum around the INFC, where the scattering
  is a tail-on collision process.
  In this paper, we will estimate the contribution of the inverse-Compton scattering process between
the cold-relativist pulsar wind and the stellar photon, since the  gamma-ray binary will provide
a unique laboratory  to investigate the  properties of the cold-relativistic pulsar wind
at the aU distance scale from the pulsar (Cerutti et al. 2008).

  The bolometric luminosity of the high-mass companion in
  the gamma-ray binaries is of the order  $L_{*}\sim 10^{38}{\rm erg~s^{-1}}$.
  This intense stellar photon field leads to
  not only observable TeV emissions through the ICS, but also  
  substantial   $\gamma\gamma$ absorption of TeV gamma-rays.
  Because the stellar radiation field  is
  anisotropic in the emission region,
  the emissivity of the ICS and optical depth of the photon-photon pair-creation process integrated
  along the line of sight depend on the orbital phase. Such anisotropic processes will cause the observed orbital modulation
  of TeV gamma-rays (e.g. Kirk et al. 1999; Bednarek 2000; Takata et al. 2014a). The secondary pairs
  created by the primary TeV gamma-rays are also a source of  very high-energy photons through
  the ICS process.  A pair-creation cascade process may develop in the stellar wind
  region and may cause the observed TeV emissions
  from  the close binary systems (Bednarek 1997, 2007; Sierpowska \&  Bednarek 2005; Sierpowska-Bartosik
  \& Torres 2007, 2008;
  Yamaguchi \& Takahara 2010; Cerutti et al. 2010) and a periastron passage of the long-orbit
  binaries (Sierpowska-Bartosik \&  Bednarek 2008).   
The anti-correlation between the GeV and TeV light curves of LS~5039 has been interpreted by using this  cascade model (Bednarek 2007)

In the X-ray bands, the observed temporal variations along the orbit 
will be mainly caused by the variations of the physical conditions of the 
pulsar wind at the shock. With the  extremely elongated orbit of
PSR~B1259-63 and PSR~J2032+4127, the distance of the shock ($r_s$)
from the pulsar
varies by about a factor of ten along the orbit (Figure~\ref{shock}). 
Hence we can in principle use gamma-ray binaries to probe
the physical properties of the pulsar wind 
 (e.g. magnetization parameter and  bulk Lorentz factor)   
as a function of radial distance. For example,  the pulsar wind
magnetization parameter (Kennel \& Coroniti 1984) can be determined as 
\begin{equation}
\sigma(r_s)=\frac{B_1^2}{4\pi\Gamma_{PW,0}N_1m_ec^2},
\end{equation}  
where, $B_1$,  $N_1$ and $\Gamma_{PW,0}$ 
 are the magnetic field,  number density and Lorentz factor of 
 the cold-relativistic pulsar wind, respectively, just before the shock.
 We can probe how the  magnetization
 parameter changes with the radial distance from 
the pulsar by examining the observed temporal variations 
(Takata \& Taam 2009; Kong et al. 2011).  
Since PSR~J2032+4127 is now approaching to the periastron passage 
 that will occur in late 2017 or early 2018, and since 
 the separation between PSR~J2032+4127 and MT91~213 changes 
 by about a factor of ten during $\sim -$1year and $\sim +1$year from
 the periastron (Figures~\ref{orbit} and~\ref{shock}),
 we can expect that the  change of the shock distance 
from the pulsar will cause a  large variation in 
the observed emission from this system, as  indicated by 
the results of the recent X-ray observations done by the Swift
(section~\ref{observations}, see also Ho et al. 2016).
Moreover, the ICS process of the cold-relativistic pulsar wind before the shock
may  boost the stellar photons from optical  to gamma-ray bands.
The next periastron passage may provide u with s a unique opportunity to
constrain the Lorentz factor $\Gamma_{PW,0}$ of the pulsar wind. 

This paper is organized  as follows.  In section~\ref{observations},
we will analyze  the  data taken from 
the Swift and Fermi-LAT.  We will confirm that the  X-ray emission observed
in 2016 is about factor of ten larger than that from   previous observations,
while the GeV flux does not show  any significant change. 
In section~\ref{model}, we describe our emission models from the
intra-binary shock and cold-relativistic pulsar wind. 
In section~\ref{result}, we will discuss the high-energy emission
within the framework of the pulsar wind/Be wind, and probe the evolution
of the magnetization parameter with a radial distance by fitting the Swift data.
In section~\ref{discussion}, we will discuss
the emission as a result of the pulsar wind/Be disk
interaction model.  We will also discuss
the possibility of the formation
of an accretion disk around the pulsar 
as a consequence  of the pulsar/Be disk interaction and its expected emission
in the high-energy bands.

\begin{table*}
\centering
\label{tabel}
\begin{tabular}{cccccccc}
\hline
PSR/Companion       & $P$ (s) & $L_{35}$ & $P_o$ (yrs) & $e$    & $a$ (lt-s) & $T_*$  & $R_*$\\\hline\hline
J2032+4127/MT91 213 & 0.143   & 1.7      & 25-50       & 0.96 & 9022     & 30000K & 10$R_{\odot}$ \\ \hline
B1259-63/LS2883     & 0.048   & 8        & 3.4         & 0.83 & 1296     & $\sim$30000K& $\sim 9R_{\odot}$\\ \hline
\end{tabular}
\caption{Parameters of the binary systems. $P$: Spin period of the pulsars. $L_{35}$: Spin down luminosity
  in units of $10^{35} {\rm erg~cm^{-2}s^{-1}}$. $P_o$: Orbital period. $e$: Eccentricity. $a$: Projected semi major
  axis. $T_*$: Surface temperature of companion star. $R_*$: Radius of B star.
  $T_*$ and $R_*$ of MT91~213 represent the values used in the calculation.
 References: Ho et al. (2016) for J2032+3127 and Negueruela et al. (2011) for B1259-63. }
\end{table*}

\section{Data analysis of  Swift and Fermi observations}
\label{observations}
\begin{figure}
\centering
\epsscale{1.0}
\plotone{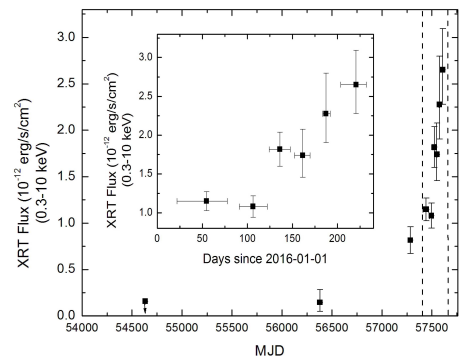}
\caption{Main figure: The long term Swift/XRT flux evolution. Inset: The flux evolution in 2016, showing an increase in X-ray flux. Details on how to obtain the data are described in section~\ref{observations}}
\label{X-ray_lc}
 \end{figure}

\begin{figure}
\centering
\epsscale{1.0}
\plotone{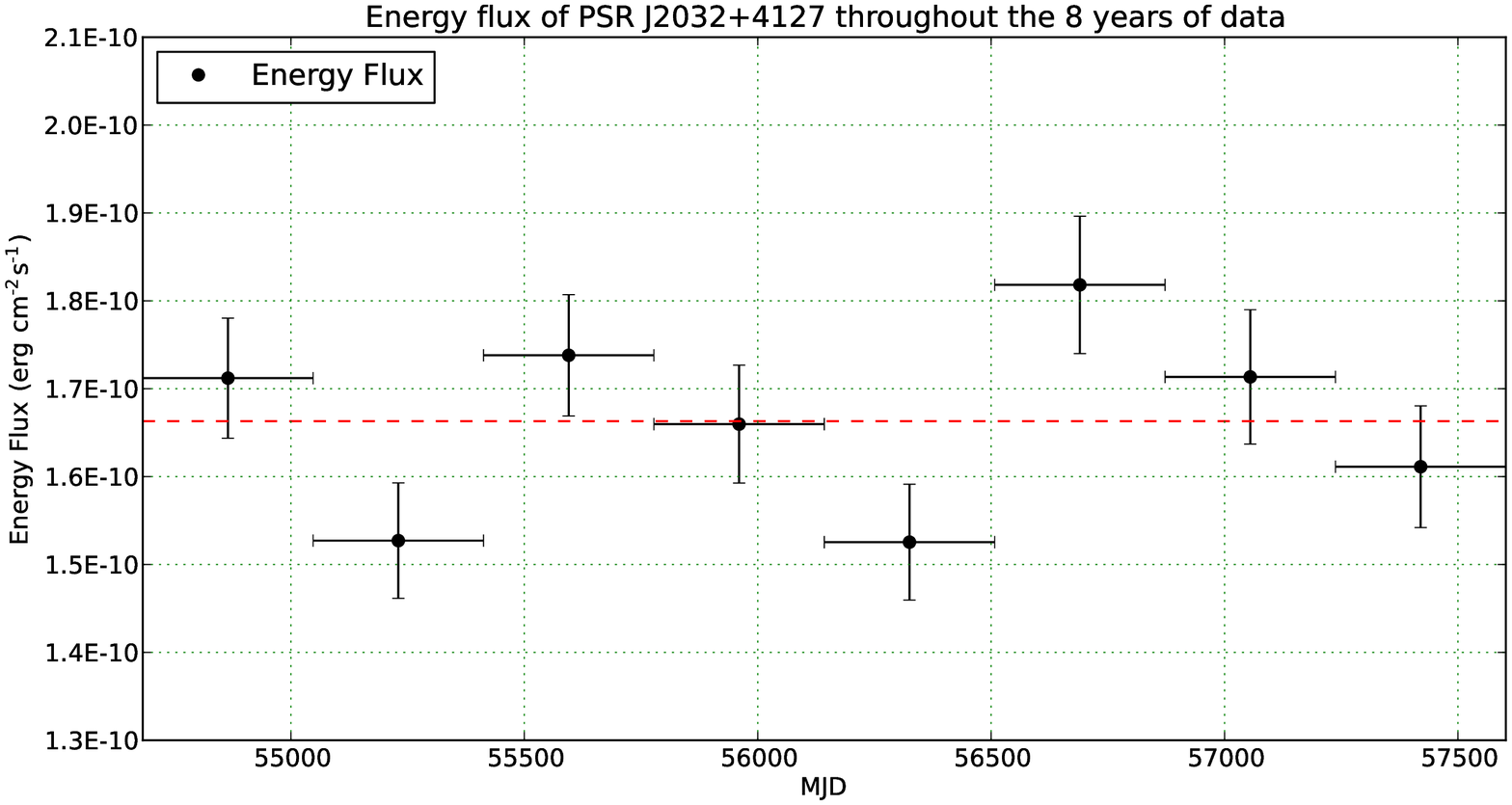}
\caption{The long term Fermi flux evolution. The horizontal line shows the average flux over 8-year observations, $F_{\gamma}=1.65\pm 0.07\times 10^{-10} {\rm erg~cm^{-2}s^{-1}}$.  }
\label{3FGL}
\end{figure}

Recently, Ho et al. (2016) reported a rapid increase in  the X-ray flux observed by Swift,
as the pulsar approached  to the periastron. In this paper,
we analyze the Fermi-LAT data as well as Swift data.
The Swift-X-ray telescope (XRT) light curve was obtained using products extracted from the XRT repository  \url{http://www.swift.ac.uk/user\_objects/} (see also Evans et al. 2007, 2009). We used a conversion factor of
1.27$\times$10$^{-10}{\rm erg~cm^{-2}ct^{-1}}$ to convert the count rates into unabsorbed fluxes,
where an absorbed power-law with $\Gamma_X=$2.37 and $n_\mathrm{H}=1.4\times 10^{22}{\rm cm^{-2}}$,
based on total galactic HI column density in the direction toward the pulsar, is assumed (Kalberla et al. 2005; Dickey \& Lockman, 1990). Since Ho et al. (2016) adopted  the
$n_\mathrm{H}=7.7\times 10^{21}{\rm cm^{-2}}$ based on the optical extinction of the companion, our unabsorbed X-ray flux is slightly higher than  that 
presented by those authors. This difference does not affect the  conclusions of the
model calculations discussed in this paper.  Due to small photon statistics, we did not include
observations shorter than 10 ks before 2014. An observation taken in 2006 had an instrumental issue and was
not included as well. An X-ray light curve is shown in Fig.~\ref{X-ray_lc}. Following up on the initial flux
increase observed in early 2016, we asked for additional dedicated XRT ToO observations since March 2016.
Three to four observations were grouped together for better signal-to-noise ratios. It is clear that the
trend in 2016 is the flux increase in X-rays.

We analyzed the gamma-ray long-term light curve of PSR~J2032+4127 using the data from the \textit{Fermi}-LAT. Photon events with energy ranging from 100 MeV to 500 GeV and time ranged from 2008-08-04 to 2016-08-02 were selected. The event class is "Pass 8 Source" and the corresponding instrumental response function is "P8R2\_SOURCE\_V6". The region of interest (ROI) is a $20^\circ \times 20^\circ$ square centered at the epoch J2000 position of the source: $(\textrm{R.A.},\textrm{Dec})=(20^{\textrm{h}} 32^{\textrm{m}} 14.35^{\textrm{s}},+41^\circ 26^\prime 48.8^{\prime\prime})$. To avoid contamination from the Earth's albedo, we excluded the time intervals when the ROI was observed at a zenith angle greater than $90^\circ$. The data reduction in this study was performed using the \textit{Fermi} Science Tools package version v10r0p5\footnote{Available at \url{http://fermi.gsfc.nasa.gov/ssc/data/analysis/software/}}. 

We first modeled the average emission from the background sources over the whole time span. The \textit{gtlike} tool was used to perform a maximum binned likelihood analysis. The source model includes all 3FGL catalog sources (gll\_psc\_v16.fit) (Acero et al. 2015) that are within $25^\circ$ from the center of the ROI, the galactic diffuse emission (gll\_iem\_v06) and the isotropic diffuse emission (iso\_P8R2\_SOURCE\_V6\_v06), available from the \textit{Fermi} Science Support Center (FSSC)\footnote{\url{http://fermi.gsfc.nasa.gov/ssc/}}. The spectral parameters for sources that are non-variable and located $5^\circ$ away from the center are fixed to their catalog values. Four extended sources within the region, Gamma Cygni, Cygnus Cocoon, HB 21, and Cygnus Loop, were modeled by the extended source templates provided by the FSSC. PSR~J2032+4127 is named as 3FGL~J2032.2+4126 in the 3FGL catalog and is modeled by a power-law with simple exponential cutoff: 
\begin{equation}
    \frac{\textrm{d}N}{\textrm{d}E}=N_0 \left(\frac{E}{E_0}\right)^{-\Gamma} \textrm{exp} \left(-\frac{E}{E_C}\right),
    \label{equation:plsec}
\end{equation}
where $N_0$ is the normalization constant, $E_0$ is the scale factor of energy, $\Gamma$ is the spectral power-law index and $E_C$ is the cut-off energy. From the binned likelihood analysis, the best-fit parameters of PSR~J2032+4127 are $N_0=(1.70 \pm 0.05)\times 10^{-11}{\rm ph~erg^{-1}cm^{-2}s^{-1}}$,
$\Gamma=-1.43 \pm 0.03$ and $E_C=4500 \pm 259$MeV. 

After fixing the spectral indices of all sources to the global fit and only leaving the normalization parameters free, the source model is then used to obtain the long-term light curve. The whole data set is then binned into eight bins, each bin is 365days. The energy range of the photons is further limited to 100 MeV-100 GeV. Local fit is performed by the \textit{gtlike} tool (binned likelihood analysis) to obtain the energy flux of PSR~J2032+4127 in each segment.
Figure~\ref{3FGL} shows the long term evolution of the flux measured by
the Fermi. No clear indication of the change in the flux  can be seen
and the average flux over eight-year observations is $F_{\gamma}=1.65\pm 0.07\times 10^{-10} {\rm erg~cm^{-2}s^{-1}}$, which will be
dominated by the pulsed emissions from the magnetosphere.

In the 1990s, the CGRO/EGRET observed the region of PSR J2032+4127 for a couple of times, including a more intensive observation during 1994 May to July. There is one source in the revised EGRET catalog (Casandjian \& Grenier, 2008), EGR J2033+4117, whose position is 0.28 deg away from 3FGL J2032.2+4126 (which has been identified as h PSR J2032+4127). The 95\% error radius of EGR J2033+4117 is 0.22 degree. Due to the slight offset,
we cannnot associate the two gamma-ray sources for certain.

It is still possible to constrain the gamma-ray emission from PSR J2032+4127 using these EGRET data. The photon flux of EGR J2033+4117 is steady over the several measurements, and is consistent with (5-6)$\times 10^{-7}$ph cm$^{-2}$~s$^{-1}$ (above 100 MeV) for all viewing periods. Due to the poorer angular resolution of EGRET compared to LAT and allowing for plausible contamination from nearby sources like Cyg X-3, this flux level may be regarded as an upper limit of any source at the position of PSR J2032+4127 over the 1990s, and is above the average LAT flux $\sim$1.4$\times 10^{-7}$ ph cm$^{-2}$~s$^{-1}$ in our analysis.

\section{Theoretical Model}
\label{model}
The origin of the high-energy TeV emission from the PSR~B1259-63/LS~2883
 system is likely related to the interaction of the pulsar wind of PSR~B1259-63 
and the Be wind/disk of LS~2883 (Tavani \& Arons 1997). 
Their interaction results in the formation of a shock, where the pulsar 
wind ram pressure and  stellar wind/disk ram  pressure are
in balance. We apply  this scenario to the PSR~J2032+4129/MT91~213 binary. 
In this section we describe our calculation methods.
We define  the capital letter $R$ and small
letter $r$ as the distances from the B star and from the  pulsar, respectively.

\subsection{Pulsar wind/stellar wind interaction}
\begin{figure*}
\centering
\epsscale{1.0}
\plotone{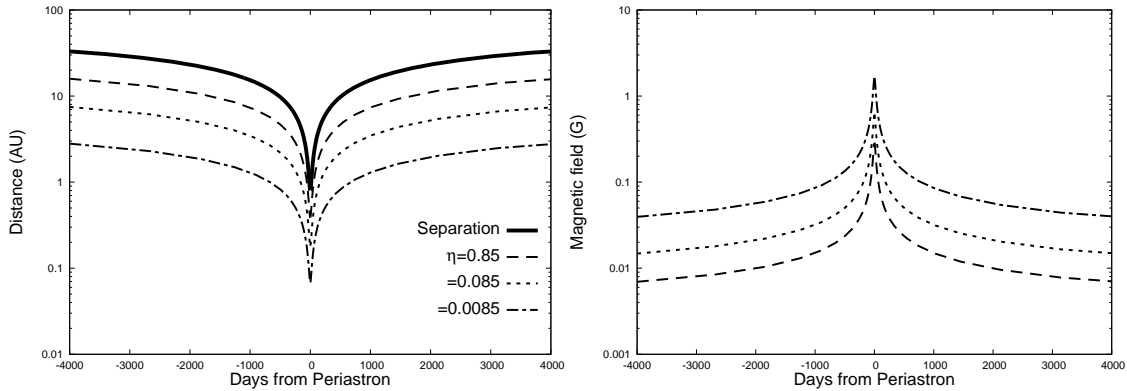}
\caption{Left: Separation between the pulsar and B star (solid line), and 
  the shock distance from the pulsars along the orbit between
  -4000days and +4000days from the periastron. The dashed ($\eta=0.85$),
 the dotted ($\eta=0.085$) and the dashed-dotted ($\eta=0.0085$) correspond 
to the cases for 
$\dot{M}_{w}=10^{-9}M_{\odot}{\rm yr^{-1}}$, $10^{-8}M_{\odot}{\rm yr^{-1}}$
 and $10^{-7}M_{\odot}{\rm yr^{-1}}$, respectively, with 
 equation~(\ref{eta}) and $v_{w}=10^8{\rm cm~s^{-1}}$. Right: The magnetic 
field strength at the periastron at the shock apex for $\eta=0.85$
 (dashed line), 0.085 (dotted line) and 0.0085 (dashed dotted line), 
respectively.  The magnetization parameter at the shock is assumed to be $\sigma(r_s)=0.1$.}
\label{shock}
 \end{figure*}

The relativistic pulsar wind, which may have  a bulk Lorentz factor
of $\Gamma_{PW,0}\sim 10^{3-6}$, interacts with the stellar wind and forms a 
cone-shaped shock separating  between the pulsar  and the companion star. The radial distance to and its opening angle of the cone-shaped shock are
determined by the ratio of the two winds (e.g. Canto et al. 1996 and references
therein), 
\begin{equation}
\eta\equiv\frac{L_{sd}}{\dot{M}_{w}v_wc},
\end{equation}
where we assume that   the pulsar wind carries the spin down-power $L_{sd}$ of 
the pulsar, $\dot{M}_w$ and $v_{w}$ is the mass loss rate and wind velocity 
of the companion star, respectively. The wind velocity of the 
stellar wind at a distance 
$R$ from the stellar surface  may be expressed  by (Waters et al. 1988) 
\begin{equation}
v_w(R)=v_0+(v_{\infty}-v_0)(1-R_*/R). 
\end{equation}
where $v_{0}\sim \sqrt{3kT_s/m_p}\sim 28$km with $T_s\sim 30,000$K being 
the temperature of the B star. The terminal velocity is estimated from 
\[
v_{\infty}\sim \sqrt{\frac{2GM_{*}}{R_*}}
  \sim 8\times 10^{7}{\rm cm~s^{-1}} \left(\frac{M_{*}}{15M_{\odot}}\right)^{1/2}
  \left(\frac{R_*}{10R_{\odot}}\right)^{-1/2}
\]
with $M_{*}$ and $R_{*}$ being the mass and radius of the B star, respectively.
Since the shock is far away from the  surface of the B star, we may assume 
 $v_{w}(R_s)=v_{\infty}$. 

 With $L_{sd}=1.7 \times 10^{35}{\rm erg~s^{-1}}$ 
of this pulsar (Lyne  et al. 2015) and  a typical mass loss rate 
$\dot{M}_w\sim 10^{-9}-10^{-7}M_{\odot}{\rm yr^{-1}}$ of an O-type or B-type  star
(e.g. Snow 1981; Smith 2006), 
the momentum ratio of this system  will be 
\begin{eqnarray}
\eta &\sim & 0.085\left(\frac{L_{sd}}{1.7\cdot 10^{35}\rm {erg~s^{-1}}}\right)
\left(\frac{\dot{M}_w}{10^{-8}M_{\odot}{\rm yr^{-1}}}\right)^{-1 } \nonumber \\
&\times&\left(\frac{v_w}{10^{-8}{\rm cm~s^{-1}}}\right)^{-1},
\label{eta}
\end{eqnarray}
indicating the stellar wind is stronger than the pulsar wind with 
a reasonable mass loss rate of the stellar wind. The distance 
to the apex of the shock cone, $r_s$, and the opening angle 
of the shock, $\theta_s$, measured by the pulsar 
 are calculated  from  (Canto et al. 1996)
\begin{equation}
r_s(\phi)=\frac{\eta^{1/2}}{1+\eta^{1/2}}a(\phi)
\label{rshock}
\end{equation}
and 
\begin{equation}
\theta_s(\eta)-\tan\theta_s(\eta)=\frac{\pi}{1-\eta},
\label{thetas}
\end{equation}
respectively,  where $\phi$ represents the orbital phase and $a(\phi)$ 
is the separation between two stars.  Figure~\ref{shock}
 summarizes the evolution of the distance (left panel) and the magnetic
 field strength (right panel) at the  shock apex along the orbit between
 -4000days and +4000days from the periastron (we did not plot the
 results for the  other orbital phase, since we are interested in the emission
 processes around  the periastron).
 In the figure, we used $\sigma(r_s)=0.1$ to calculate the magnetic field
 strength. As the figure shows, the shock distance
 and the magnetic field at the shock 
change by  more than factor of ten along the orbit. Since the pulsar orbit 
velocity ($v_p\sim 10^{6-7}{\rm cm~s^{-1}}$) 
is at least  about a factor of ten slower  than the stellar wind velocity 
($v_w \sim 10^8{\rm cm~s^{-1}}$), we assume that the shape of the shock cone is axially symmetric
about   the axis connecting the  two stars.

\subsection{Pulsar/Be disk interaction}
\label{disk}
\begin{figure}
\centering
\includegraphics[width=0.5\textwidth]{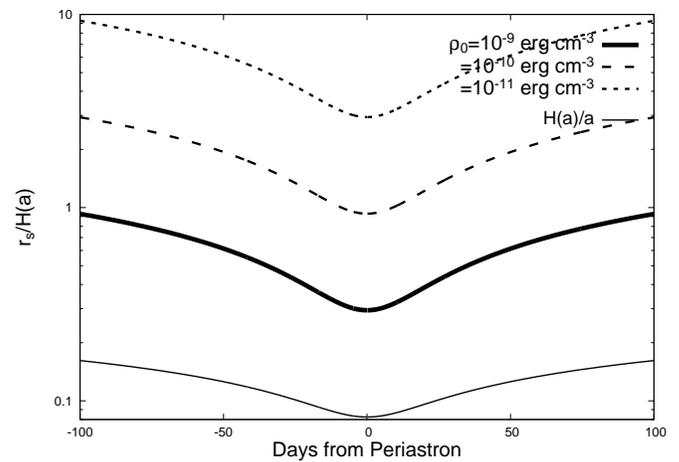}
\caption{Shock distance [equation~(\ref{shockdisk})] for
  the pulsar/Be disk  interaction, if the pulsar interacts with the Be disk at 
the given orbit phase (x-axis). 
The disk structure is calculated with $n=3.5$, $\beta=1.5$, 
$H_0=0.02R_*$ and circular velocity $v_{d,K}=500{\rm km~s^{-1}}(R_*/R)^{1/2}$. The  
relative velocity is assumed and is calculated
 from $v_r\sim \sqrt{v_p^2+v_{d,K}^2}$. The solid, dashed and dotted lines 
show the ratio of the shock distance and disk scale height at the pulsar 
 for the base density of $\rho_0=10^{-9}{\rm g~cm^{-3}}$,
 $10^{-10}{\rm g~cm^{-3}}$ and $10^{-11}{\rm g~cm^{-3}}$, respectively. The 
dashed-dotted line shows the ratio of the scale height and separation, 
$H/a$.  A cavity of the pulsar wind around the pulsar may 
 be formed when $r_s/H$ is less than unity.
}
\label{bow}
 \end{figure}
\begin{figure}
\centering
\includegraphics[width=0.5\textwidth]{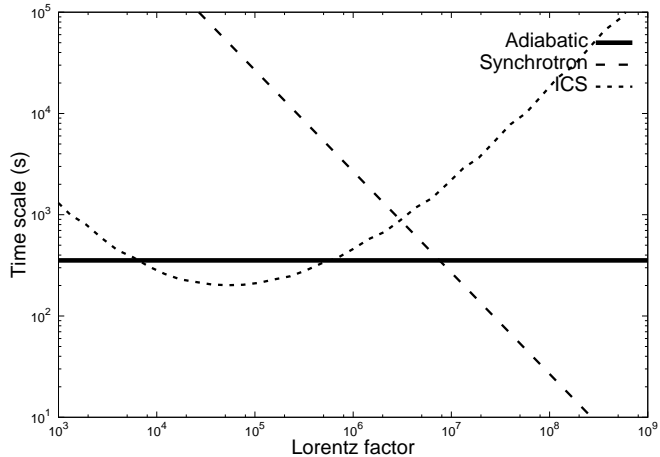}
\caption{Cooling time scale as a function of the Lorentz factor at periastron 
 and at the apex of the shock cone with $\eta=0.085$. 
The magnetization parameter is $\sigma_0=0.1$ ($B_2\sim 0.3$G).  
The solid, dashed and dotted lines are time scale of 
the adiabatic cooling, synchrotron cooling and ICS cooling, respectively. 
}
\label{cool}
 \end{figure}

Since  the Be-type star, MT91~213, forms a dense equatorial disk outflow,  
the pulsar may  interact with the Be disk at the periastron passage, as in the case with the gamma-ray binary PSR~B1259-63/LS2883 system.  
The decretion disk model has been explored  to describe the structure of the 
Be disk (Lee et al. 1991;  Carciofi \& Bjorkman 2006; Okazaki et al. 2011),
 and it implies the disk mass 
density ($\rho_d$)  and the scale height  ($H$)  are described by 
\begin{equation}
\rho_d(R,z)=\rho_0\left(\frac{R_*}{R}\right)^n{\rm exp}\left(-\frac{z^2}{2H^2}\right),
\end{equation}
and
\begin{equation}
H(R)=H_0\left(\frac{R}{R_*}\right)^{\beta},
\end{equation}
respectively, where $n\sim 3-3.5$ and $\beta\sim 1-1.5$ 
($n=3.5$ and $\beta=1.5$ for an  isothermal disk). In addition, we apply  
$H_0\sim 0.02R_*$ and the base density
$\rho_0\sim 10^{-11}-10^{-9}{\rm g~cm^{-3}}$. 

Takata et al. (2012) studied  the orbital modulations of
the X-ray/TeV emission from  PSR~B1259-63/LS2883 system.
They  argued that with 
the  larger base density ($\rho_0\sim 10^{-9}{\rm g~cm^{-3}}$) for the Be disk,
 the pulsar wind creates a cavity in the disk gas  and 
this causes a significant increase in the conversion efficiency from pulsar
 spin-down power to the shock-accelerated particle energy. This explains  
the double-peak structure of the X-ray light curves of the gamma-ray binary 
PSR B1258-63/LS 2883. The pulsar/Be disk interaction with a
 smaller base density ($\rho_0< 10^{-10}{\rm g~cm^{-3}}$) causes 
 no cavity in  the pulsar  wind  and hence no enhancement in the emission. 

With the elongated orbit and the long orbital period 
of the PSR~J2032+4127, a strong pulsar/Be disk interaction will be possible at
 only periastron passage.  The radius of the shock from the pulsar 
may be obtained from a pressure balance condition, 
\begin{equation}
r_s=\left(\frac{L_{sd}}{2\pi \rho_dv^2_r c}\right)^{1/2},
\label{shockdisk}
\end{equation}
where $v_r$ is the relative velocity between the pulsar and the disk rotation. 
If the pulsar interacts with the Be disk at the periastron, where the 
separation of the two stars is $a\sim 1$au,  the mass density
 and scale height of the disk at the pulsar 
are  $\rho(a)\sim 2\times 10^{-5}\rho_0$ and 
$H(a)\sim 10^{12}{\rm cm}$, where we used $n=3.5$, $\beta=1.5$, 
$H_0=0.02R_*$ and $R_*=10R_{\odot}$. The orbital  velocity of the pulsar and 
circular velocity of the disk are $v_{p}\sim 10^{7}{\rm cm~s^{-1}}$ 
and $v_{d,K}(a)=v_{d,K}(R_*)(R_*/1{\rm AU})^{1/2}\sim 10^{7}{\rm cm~s^{-1}}$, respectively, where we used $v_{d,K}(R_*)=5\times 10^{7}{\rm cm~s^{-1}}$.  The
radial velocity of the disk matter, $v_{d,r}\sim
0.1c_s\sim 10^5{\rm cm~s^{-1}}$ with $c_s\sim 10{\rm km~s^{-1}}$
being the sound speed (Okazaki et al. 2011),  is much slower than
the pulsar motion and the disk's circular motion.
Hence, by assuming the relative velocity $v_{r}\sim \sqrt{v_p^2+v_{d,K}^2}$, 
the shock distance calculated from  the equation~(\ref{shockdisk}) is  
$r_s\sim 10^{12}{\rm cm}(\rho_0/10^{-10}{\rm g~cm^{-3}})^{-1/2}$ at
the periastron.  A pulsar/Be disk interaction
may cause  a cavity in  the pulsar wind around 
the pulsar,  provided that $r_s/H(a)\le 1$,
which yields  $\rho_{0}\ge 10^{-10}{\rm g~cm^{-3}}$.

Figure~\ref{bow}  summarizes the ratio of the shock distance given by 
equation~(\ref{shockdisk}) and the scale height of the disk,
provided that the pulsar interacts with the disk
at the given orbital phase in the horizontal axis.
The figure summarizes  the shock distance for the epoch between   -100days and 
+100days from the periastron. 
 The different lines show the case for the base density 
of $\rho_0=10^{-9}{\rm erg~cm^{-3}}$ (solid line), 
$10^{-10}{\rm erg~cm^{-3}}$ (dashed line) and 
$10^{-11}{\rm erg~cm^{-3}}$ (dotted line), respectively.
We find from  the figure that a  cavity would form ($r_s/H<1$) 
 at the periastron passage if $\rho_0>10^{-10}{\rm erg~cm^{-3}}$.
 If the base density is less than $\rho_0<10^{-10}{\rm erg~cm^{-3}}$, on 
the other hand,  the pulsar wind will  strip
off an outer part of the Be disk, truncating it  at a radius smaller 
than the pulsar orbit, and the pulsar wind/Be disk interactions will not affect to
the observed emission. 

\subsection{Shock emissions}
We calculate the emissions from the shock due to the interaction between 
the pulsar wind and stellar wind, as follows.
The magnetic field at the shock radius but
before the shock is calculated from 
\begin{equation}
B_1=\sqrt{\frac{{L_{sd}}\sigma_0}{r_s^2c(1+\sigma_0)}},
\end{equation}
where we defined $\sigma_0\equiv \sigma(r_s)$.
At the shock, the kinetic energy 
of the pulsar wind is converted into the internal energy of the shocked 
pulsar wind. Applying  a  jump condition of a perpendicular MHD shock, we 
calculate the  velocity $v_{pw,2}$,  the magnetic field $B_2$, 
 and the gas pressure $P_2$ 
of the shocked pulsar wind at the shock (Kennel \& Coroniti 1984).  
 For the particle kinetic energy dominating the  un-shocked flow, that is, 
for the low $\sigma_0$ regime, we obtain  
\begin{equation}
v_{pw,2}\sim\frac{c}{3}\sqrt{\frac{1+9\sigma_0}{1+\sigma_0}},
\label{vpw2}
\end{equation}
\begin{equation}
B_2\sim 3B_1(1-4\sigma_0)
\end{equation}
and 
\begin{equation}
P_2\sim \frac{2(1-2\sigma_0)}{3(1+9\sigma_0)^{1/2}(1+\sigma_0)}
\frac{L_{sd}}{4\pi r_s^2c}.
\end{equation}
We assume that the post-shock  pulsar wind flows along the shock surface. 
Along the downstream flow,  we assume a conservation of the magnetic flux 
$B(r)=r_sB_2/r$. For the spherical symmetric flow,  
the bulk velocity decreases with the distance from the shock 
(Kennel \& Coroniti 1984).  However,
 the numerical simulations imply that the post-shock bulk flow for the 
binary systems does not simply decrease but increases with the 
distance from the shock, because of a rapid expansion of the flow in the 
downstream region (Bogovalov et al. 2008). Furthermore, the 
high-energy emissions take  place  in the vicinity of the shock radius.  In this 
study, therefore,  we assume the velocity field of the post-shock pulsar
 wind with $v_{pw}(r)=$constant.

 We assume that the electrons and positrons in the pulsar wind
 are accelerated by the shock and form a power law distribution over several decays in energy;
\begin{equation}
f_0(\gamma)=K_0\gamma^{-p},~~\Gamma_{PW,0}\le\gamma\le\gamma_{max}, 
\label{inidis}
\end{equation}
where we assume that 
the minimum Lorentz factor of the accelerated particles is equal to 
the average Lorentz factor of the particles forming
the cold-relativistic pulsar wind (section~\ref{icpw}). 
We assume the maximum Lorentz factor by balancing between the acceleration
time scale $\sim \gamma_{max}m_ec/(\xi eB_2)$ and the synchrotron
loss time scale $\sim 9m_e^3c^5/(4e^4B_2^2\gamma_{max})$, yielding
 $\gamma_{max}=[9\xi m_e^2c^4/(4e^3 B_2)]^{1/2}$,  where
  $\xi$ represents the efficiency of the acceleration, which will be
  $\xi\le 1$. The efficiency of the acceleration in the gamma-ray
binaries  has been  explained from the observed  spectra of
  very high-energy emissions; for example,  a  high efficiency is  
required to explain the TeV emissions of LS~5039 (Zabalza et al. 2013).  
For PSR J2032+4127/MT91 213,  
it is difficult to  discuss the efficiency, since TeV 
 emission from this source has not yet  been detected.  
In the calculations, we found that  the predicted X-ray flux and 
0.1-10TeV flux are insensitive to the efficiency if $\xi>10^{-4}$.
  In this study, therefore, we present the results with $\xi=1$. Future 
TeV observations at the periastron passage may provide an addition information constraining  the efficiency.

For the injected particles at the shock, we assume $p=2$ for the
power-law index of the distribution.
Since the particle energy density $\epsilon$ 
 is related with the pressure $P_2$ as $P_2=\epsilon/3$,  
the normalization factor $K_0$ of equation~(\ref{inidis}) is  
calculated from  the relation $P_2=m_ec^2\int_{\gamma_{min}}^{\gamma_{max}}
\gamma f_0(\gamma)d\gamma/3$.

We solve the evolution of the Lorentz factor after the shock with a simple 
one-dimensional treatment, that is, 
\begin{equation}
\frac{d\gamma_e}{dr}=\frac{1}{v_{pw}}
\left[\left(\frac{d\gamma}{dt}\right)_{ad}
+\left(\frac{d\gamma}{dt}\right)_{syn}
+\left(\frac{d\gamma}{dt}\right)_{ICS}
\right].
\end{equation}
We assume the adiabatic  loss given by
\[
\left(\frac{d\gamma}{dt}\right)_{ad}=\frac{\gamma}{3n}\frac{dn}{dt}
=-\frac{2\gamma v_{pm}}{3r},
\]
where $n\propto r^{-2}$ is the particle number density.  The synchrotron 
loss is given as 
\begin{equation}
\left(\frac{d\gamma}{dt}\right)_{syn}=-\frac{4e^2B^2\gamma^2}{9m_e^3c^5}.
\end{equation}
 We calculate  the ICS energy loss rate using  
\begin{equation}
\left(\frac{d\gamma}{dt}\right)_{ICS}
=-\int\int(E-E_s)\frac{\sigma_{ICS}c}{m_ec^2E_s}\frac{dN_s}{dE_s}
dE_sdE,
\end{equation}
where $dN_s/dE_s$ is the stellar photon field distribution and $\sigma_{ICS}$ 
is the cross-section with the isotropic photon field. For the soft-photon 
field, we consider the blackbody radiation from the B star and apply the 
Planck function with the temperature $T_s=30,000$K, which is the case for
 companion star of PSR~B1259-63 (Negueruela et al., 2011).  Figure~\ref{cool} 
summarizes the time scale of the cooling at the periastron and at 
the apex of the shock cone.

\begin{figure}
\centering
\includegraphics[width=0.5\textwidth]{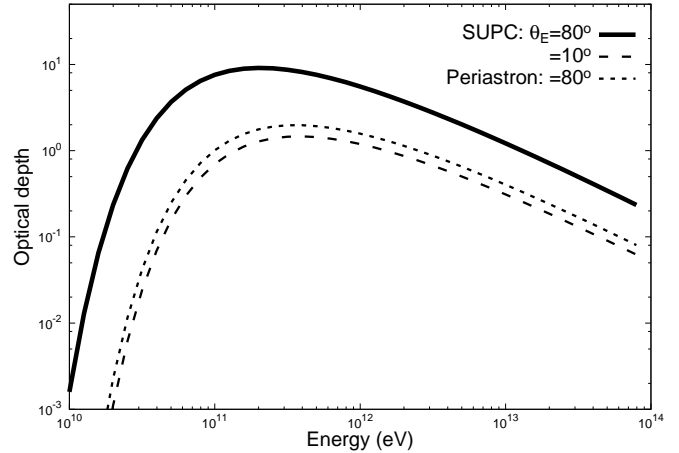}
\caption{The optical depth of the photon-photon pair-creation process
  as a function of the energy of the photon that propagating
  toward the observer. The solid and dashed lines  show the optical depth at
  the SUPC  with assumed
  system inclination angle $\theta_E=10^{\circ}$ and $80^{\circ}$, respectively.
  The dotted line is the optical depth at the periastron with $\theta_E=80^{\circ}$.
   The anomaly of the direction of Earth is $\phi_E=230^{\circ}$. }
\label{tau}
 \end{figure}

The TeV gamma-ray produced by the ICS process may be converted
into an electron-positron pair through the photon-photon pair-creation process with the stellar photon. The cross section of this  process is calculated from
\begin{equation}
  \sigma_{\gamma\gamma}(E_s,E_{\gamma})
  =\frac{3}{16}\sigma_T(1-v^2)
  \left[(3-v^4{\rm ln}\frac{1+v}{1-v}-2v(2-v^2)\right],  
  \end{equation}
where
\[
v(E_s,E_{\gamma})=\sqrt{1-\frac{2}{1-\cos\theta_{\gamma\gamma}}\frac{(m_ec^2)^2}{E_sE_{\gamma}}},
\]
where $\theta_{\gamma\gamma}$ is the collision angle between the soft photon and gamma-ray, and  it is a function
of position.  The optical depth is calculated from $\tau(E_{\gamma})=\int\int
\sigma_{\gamma\gamma}(E_s,E_{\gamma})n_s(E_s)dE_{s}d\ell$,
where $n_s(E_s)$ represents the energy  distribution of the stellar
radiation number density  (see Table 1 for the parameters of the companion star).
Figure~\ref{tau} summarizes
the optical depth of the high-energy photons emitted from  the shock apex ($\eta=0.085$);
different lines in the figure represent difference system inclination angles $\theta_E$ and
different positions of the pulsars along the orbit.  We can find in   
Figure~\ref{tau} that the optical depth  of the  $0.1-1$TeV photons emitted around
the periastron and the  SUPC exceeds unity, indicating the TeV photons emitted  during the
periastron passage are significantly absorbed by the stellar photons.

\subsection{ICS process of the cold-relativistic pulsar wind}
\label{icpw}
\begin{figure*}
\centering
\epsscale{1.0}
\plotone{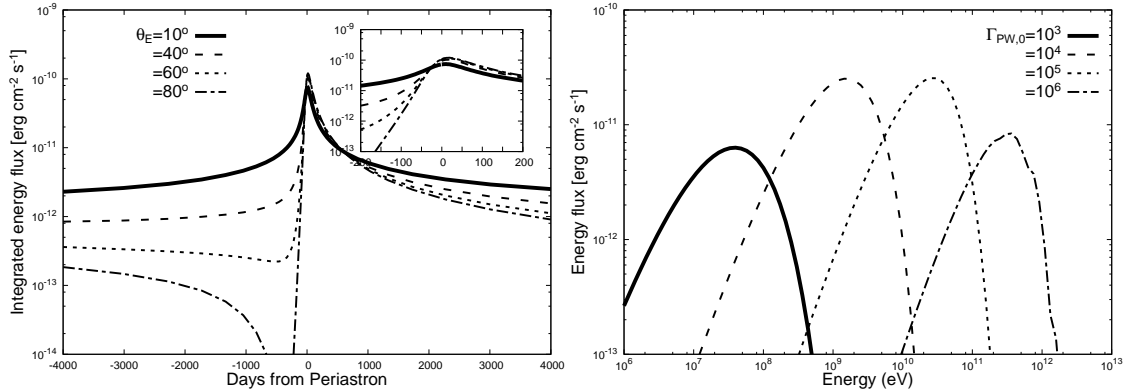}
\caption{The orbital modulations (left panel)  and 
the spectra (right panel) of the ICS 
process between the cold-relativistic pulsar wind and stellar photons. 
Left: The orbital modulations for the different 
observing angle ($\theta_E$)  measured from the direction perpendicular to 
the orbital plane.  The results are for $\Gamma_{PW,0}=10^4$.  
Right panel: the calculated spectra at the periastron.  The results are 
for  $\theta_E=60^{\circ}$. 
 }
\label{cpw}
 \end{figure*}
The pulsar binaries will be a laboratory to search for the evidence 
of the high-energy gamma-rays produced by the  ICS process of
the cold-relativistic pulsar wind (Cerutti et al. 2008).
In this study, we assume the pulsar wind is isotropic and is formed near the 
light cylinder  of the pulsar, which is $r_{lc}\sim 6.8\times 10^8$cm 
for PSR~J2032+4127. Since there is a theoretical uncertainty in  
the particle distribution of the cold-relativistic 
pulsar wind, we explore the process with a relativistic Maxwell distribution 
of  the form,
\begin{equation}
f(\Gamma_{PW})\propto \Gamma_{PW}^2{\rm exp}\left(-\frac{3\Gamma_{PW}}
{\Gamma_{PW,0}}\right),
\end{equation}
which provides an  average Lorentz factor of $\sim \Gamma_{PW,0}$. 
The normalization is calculated from the condition that
\begin{equation}
m_ec^2\int_0^{\infty}\Gamma_{PW}f(\Gamma_{PW})d\Gamma_{PW}
=\frac{L_{sd}}{4\pi r^2c (1+\sigma)}.
\end{equation}
Due to  the unknown energy conversion process from the magnetic
energy to the particle energy, the magnetization parameter will
evolve with the radial distance from the pulsar. 
From the energy conservation,  the average Lorentz factor of
the cold-relativistic pulsar wind just before the shock may be described  as 
$\Gamma_{PW,0}(r_s)=\Gamma_L(1+\sigma_{L})/[1+\sigma_0]$,
 which provides 
$\Gamma_{PW,0}(r_s)\sim \Gamma_{L}\sigma_{L}$ in the limit of $\sigma_{L}\gg 1$ 
 and $\sigma_0\ll 1$, where $\Gamma_{L}$ and $\sigma_{L}$ are the
 Lorentz factor and the magnetization parameter
 at the light cylinder, respectively. The pairs created 
inside the light cylinder lose  their momentum perpendicular to the magnetic 
field line  via synchrotron radiation,  and they will 
eventually escape from the light cylinder with a Lorentz factor 
of $\Gamma_{L}\sim 10$. Hence, the magnetization parameter at the light cylinder 
may be estimated from
\begin{equation}
\sigma_L=\frac{B_{L}^2}{4\pi\Gamma_L\kappa n_{GJ,L}m_ec^2}
\sim 9\times 10^2\left(\frac{\Gamma_L}{10}\right)^{-1}
\left(\frac{\kappa}{10^5}\right)^{-1}
\end{equation}
where $\kappa$ is the multiplicity and 
we applied  $P=0.143$s and $B_{L}=4.3\times 10^3$G of PSR~J2032+4127.  
From the observations of the pulsar wind nebulae, the multiplicity is 
expected to be   $\kappa\sim 10^{5-6}$ for Crab-like young pulsars
(Tanaka \& Takahara 2010) and $\kappa\sim 10^{2-3}$ 
for the Vela-like pulsars (Sefako \& de Jager, 2003). Theoretical studies
have implied  the multiplicity
$\kappa\sim 10^{2-5}$ (Hibschman\& Arons 2001; Timokhin \& Harding 2015).
Hence we assume the typical value of the Lorentz factor of
the pulsar wind is in the range of $\Gamma_{PW, 0}\sim \Gamma_L\sigma_L\sim 10^{3-6}$. 

The power per unit energy per unit solid angle of the ICS  process 
with an anisotropic soft-photon field  may be calculated from  
\begin{equation}
\frac{dP_{ICS}}{d\Omega}= {\cal D}_{ICS}\int (1-\beta\cos\theta_0)I_b/h
\frac{d\sigma'_{ICS}}{d\Omega'}d\Omega_*,
\end{equation} 
where $d\Omega_*$ is the solid angle of the sky covered by the B star
measured from the emission region, ${\cal D}_{ICS}=\Gamma^{-1}(1-\beta\cos\theta_1)^{-1}$ with 
$\theta_1$ (or $\theta_0$) describing the angle between
the direction of the particle motion and the propagating direction of the scattered photons (or background photons). In addition, $I_b$ is the Planck function of the 
stellar photon field, and $d\sigma'_{ICS}/d\Omega'$ is the 
differential Klein-Nishina cross section.

Figure~\ref{cpw} shows the expected light 
curves (left panel) and spectra at the periastron (right panel)  
for the different system inclination angles relative to the 
Earth viewing angle ($\theta_E$) and the average Lorentz factor of the 
pulsar wind $\Gamma_{PW,0}$, respectively.
As the left panel indicates,
the calculated  emissions for a larger viewing angle are
suppressed  at around $-150$days from
the periastron. This is because the INFC occurs  at 
$\sim -150$days from  the periastron  (Figure~\ref{system}),
and because at around the INFC the ICS process between a soft photon from 
the B star and a pulsar wind particle moving toward 
the Earth is a tail-on collision process, which reduces the ICS efficiency.
 Since the SUPC is $\sim +10$ days after periastron,
 the calculated flux measured from the Earth becomes maximum  around, 
 but after, periastron. As the right panel shows,
 the energy flux in the GeV energy 
bands is $\sim 10^{-10}{\rm erg~cm^{-2}~s^{-1}}$ for $\Gamma_{PW,0}\sim 10^{4-5}$, 
which is close to the observed flux 
level ($\sim 2\times 10^{-10}{\rm erg~cm^{-2}~s^{-1}}$) of the pulsed emission
 from  the Fermi-LAT. Hence, we expect that if $\Gamma_{PW,0}\sim 10^{4-5}$, 
the Fermi-LAT will observe an increase in the GeV flux during  
the next periastron passage.

\section{Model results}
\label{result}
In the current calculation, we applied the system parameters suggested
by Ho et al. (2016), namely, the orbital period of $P_{ob}=17,000$days,
and  the anomaly of the direction of the Earth $\phi_{E}\sim 230^{\circ}$
(Figure~\ref{system}). In addition, we apply $d=1.5$kpc (Lyne et al. 2015)
and  MJD~58069 as the periastron. We have  not present the results of the calculations for
the orbital phase $<-4000$ days and $>+4000$ days from the periastron,
since we are interested in the emission processes around the periastron.
We would like to note that within the current
uncertainties of the orbit parameters
($P_{ob}$, $e) \sim $(25years, 0.93) in Lyne et al. (2015) and (50years, 0.96)
given by Ho et al. (2016), the main conclusions of the model results
are not affected, since the system sizes predicted by two authors
are very similar  to each other. 
 
\subsection{X-ray/TeV emissions}
As described in the previous section, the Swift observed a rapid increase in
the emissions  in 0.3-10keV energy bands
after $\sim$MJD56250 (see Figure~\ref{X-ray_lc}).
These X-ray emissions are probably caused as a result
of the pulsar wind/stellar wind interaction, since the pulsar/Be disk
interaction will be important only at around the periastron (say, between
-100 days and +100 days), as discussed in the section~\ref{disk}.
In this section, therefore,  we will discuss the X-ray and TeV emissions from
the shock caused by the pulsar wind and stellar wind interaction, and will
examine the pulsar wind properties by comparing the model results with
the observed  X-ray light curve. 
\begin{figure*}
  \centering
  \epsscale{1.0}
  \plotone{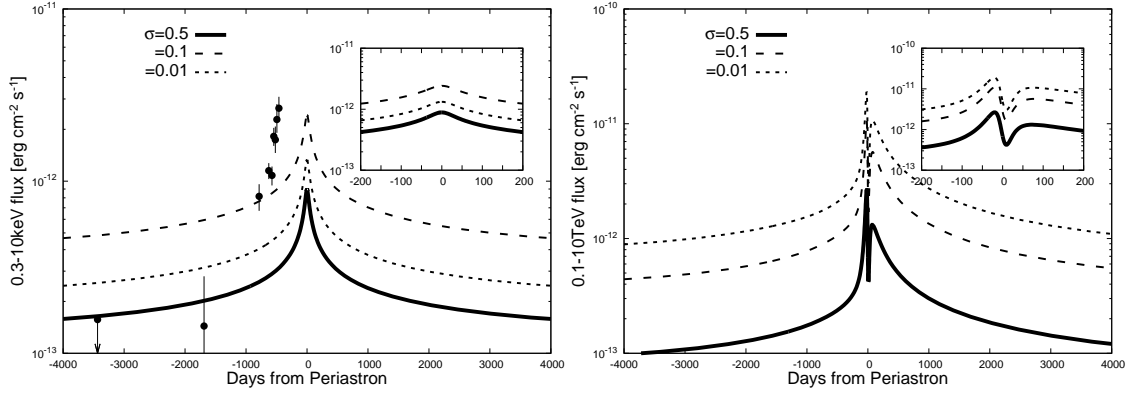}
\caption{Modulation in 0.3-10keV energy bands (left panel) and 
in $>$1TeV energy bands (right panel)  with constant magnetization 
parameter. The solid, dashed and dotted lines are results for $\sigma_0=0.5$,
 0.1 and 0.01,respectively. The sub-panel enlarges the modulation between 
-100days and 100days from the periastron.  The results are for 
$\Gamma_{PW,0}=10^4$, $\eta=0.085$ and $\theta_E=60^{\circ}$.
}
\label{light}
 \end{figure*}
\begin{figure*}
  \centering
  \epsscale{1.0}
  \plotone{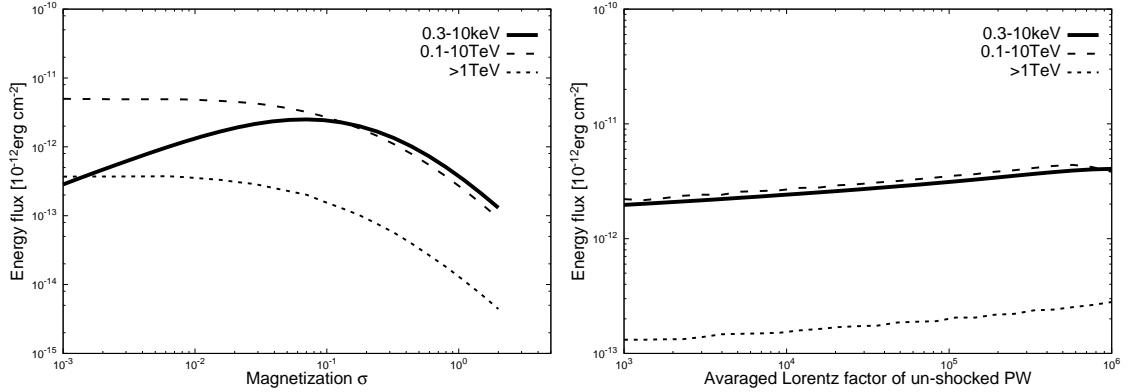}
\caption{The dependency of the calculated X-ray (solid line) and 
TeV (dashed line) as a function of the magnetization parameter with $\Gamma_{PW,0}=10^4$  (left panel) 
and Lorentz factor (right panel) with $\sigma_0=0.1$. 
The results are for  the periastron  and $\theta_E=60^{\circ}$.}
\label{depend}
 \end{figure*}
\begin{figure}
  \centering
  \epsscale{1.0}
  \plotone{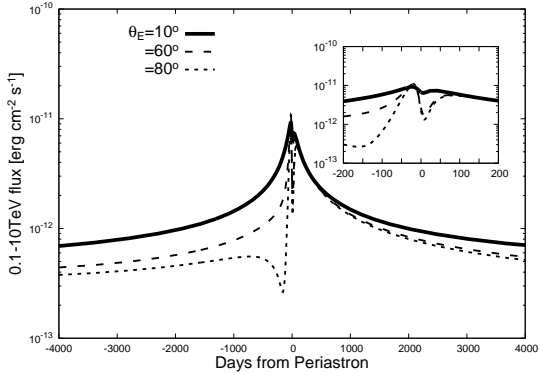}
  \caption{The dependency of the calculated 0.1-10TeV light curve
    on the Earth viewing angle measured from the direction perpendicular to
   the  orbit plane. The solid, dashed and dotted lines
    are for $\theta_E=10^{\circ}$, $60^{\circ}$ and $80^{\circ}$,
  respectively. In addition, $\sigma_0=0.1$ and $\Gamma_{PW,0}=10^4$. }
  \label{tevlight}
\end{figure}

\subsubsection{Constant $\sigma$}
\label{constant}
First, we consider the case where  the  magnetization parameter is constant with
the radial distance from the pulsar, namely, 
the magnetization parameter at the shock 
does not depend on  the shock distance. We also ignore the Doppler boosting effect
caused by the finite velocity of the shocked pulsar wind, which will be discussed in section~\ref{dop}. Figure~\ref{light} summarizes
the calculated flux and orbital modulation in the  0.3-10keV energy bands  (left 
panel) and in the $0.1-10$TeV energy bands (right panel); the solid, dashed,  
and dotted lines are for $\sigma_0=0.5$,  0.1, and 0.01, respectively. In the calculation,  we assumed $\eta=0.085$ (that is, $\dot{M}_W\sim 10^{-8}M_{\odot}{\rm yr^{-1}}$), and $\Gamma_{PW,0}=10^4$. 
 In the left panel of the figure,  the results of the Swift observations
 are also displayed.

 We find in Figure~\ref{light}
 that the calculated light curve with constant magnetization parameter
 predicts an   increase in the X-ray flux slower than that of
 the Swift observations after $\sim -2000$ days from the periastron.
 The synchrotron luminosity is
  roughly proportional to $L_{syn}\propto r_sB^{2}(r_s)\propto  1/r_s$
  for a constant magnetization parameter.  As Figure~\ref{shock} shows,
  the shock distance decreases by about a factor of $\sim 2$ during $\sim
  -2000$days and $\sim -1000$days, and therefore,
  the calculated flux slightly  increases during that epoch, which cannot
  explain the increase in the flux (a factor of  $\sim 5-10$) measured 
  by the Swift.

  Figure~\ref{depend} shows how the calculated flux depends on 
the model parameters, namely, the  magnetization  parameter (left panel) 
and the  average Lorentz factor of the
 un-shocked  pulsar  wind $\Gamma_{PW,0}$ (right panel).   
 As seen in the left panel, 
 the calculated X-ray flux (solid line) reaches its maximum value at 
 the magnetization parameter $\sigma_0\sim 0.1$.
 This is because the internal energy ($\propto P_2$)
 of the post-shock flow at the shock given by the jump condition 
 decreases as the assumed 
magnetization parameter increases (e.g. figure~4 of Kennel \& Coronit 1984), 
while the magnetic field at the shock ($B_2$) increases as the magnetization 
parameter increases. The former tends to decrease the calculated X-ray flux,
 while the latter increases it. These two 
effects compensate each other and  the calculated X-ray flux becomes 
maximum at the magnetization parameter $\sigma_0\sim 0.1$.
We find therefore that it is difficult to explain the observed X-ray flux
of Swift after  $\sim -1000$days with using the magnetization
parameter $\sigma_0\sim 1$.  By comparing
the calculated flux and Swift observations, therefore,
we conclude that the magnetization parameters at the shock  is of order of $\sigma_0\sim 0.1$ at $\sim -1000$days. It is  possible that the Lorentz factor $\Gamma_{PW,0}$ of the cold-relativistic pulsar
wind evolves with the shock distance from the pulsar.  In the right panel of Figure~\ref{light},
however, we can see that the calculated fluxes in X-ray/TeV energy bands are less dependent on the
minimum Lorentz factor of the shocked particles at the shock ($\Gamma_{PW,0}$).

In the calculated  TeV light curve (right panel in Figure~\ref{light}),
  we can see an asymmetry relative to the 
  periastron. Since the TeV photons from the shock are
  produced by the ICS process, the asymmetry  is introduced  by
  the dependence of the  collision angle between the stellar photons and
  the shocked pulsar wind particles that emit the photons toward the Earth,
  as in the case with the orbital modulation
  of the ICS of the cold-relativistic pulsar wind (Figure~\ref{cpw}).

  For the TeV photons, the optical depth of the
  pair-creation process  depends on the orbital phase, and it becomes
  the maximum at the SUPC for the photons
  traveling toward the Earth. This effect can be seen as a  rapid drop
  in the calculated flux around $\sim +10$days.
  The calculated TeV emission from the shock depends on the Earth viewing
  angle, as summarized in Figure~\ref{tevlight}. 

  \subsubsection{Doppler boosting effect}
  \label{dop}
 \begin{figure}
  \centering
  \epsscale{1.0}
  \plotone{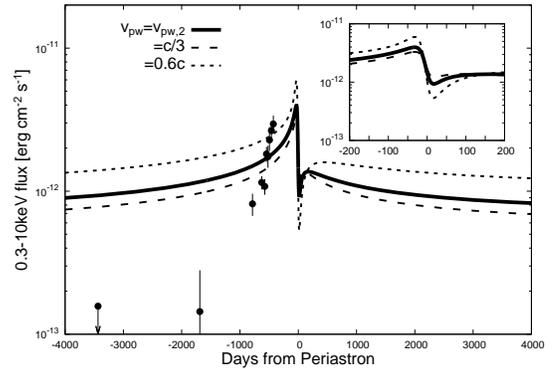}
  \caption{Orbital modulation with Doppler boosting effect due to the finite velocity
    of the shocked pulsar wind. The different lines assume the different
    flow velocity; $v_{pw}=v_{pw,2}$  (solid line) given by the jump condition,
    $c/3$ (dashed line) and $0.6c$ (dotted line), respectively.
    The assumed inclination
    angle of the system is $\theta_E=60^{\circ}$. The results are for
    $\eta=0.085$, $\Gamma_{PW,0}=10^4$, and $\sigma_0=0.1$. }
  \label{light-d}
\end{figure}

  In the previous sections, we ignored the effect
  of the motion of the post-shocked pulsar wind. In such a case,
  as Figure~\ref{light} indicates, the orbital modulation of the X-ray emission is symmetric  relative to
  the periastron. Asymmetry in the X-ray light curve
  as well as in the TeV light curve will be introduced by  the finite
  velocity of the shocked pulsar wind.

  The finite velocity of the shocked pulsar wind will cause
  the Doppler boosting effect, which enhances or suppresses  the observed emissions.
  It has been suggested that the observed
  modulation of the X-ray emissions  from gamma-ray
  binary LS~5039 is a result  of  the Doppler boosting effect
  (Dubus et al. 2010; Takata et al 2014a). The Doppler factor  is calculated from 
  \begin{equation}
    {\cal D}=\frac{1}{\Gamma_{pw}\sqrt{1-(v_{pw}/c)\cos\theta_{pw}}},
    \label{doppler}
  \end{equation}
  where $\Gamma_{pw}$ is the Lorentz factor of the bulk motion of
  the post-shocked flow, and $\theta_{pw}$ is the angle of the flow direction
  and the line of sight. We assume that the velocity of the post-shocked
  pulsar wind flow is constant.
  
  Figure~\ref{light-d} summarizes the Doppler boosting effect with the different
  velocities of the post-shocked flow,
  $v_{pw}=v_{pw,2}$ (solid lines) given by the jump condition,
  $c/3$ (dashed line) in the limit of $\sigma_0=0$ and $0.6c$ (dotted line).
  We can see in the figure that the Doppler boosting effect enhances
  the emissions before the periastron, while suppressing  after it.
  With $\eta< 1$, namely for the stellar wind stronger than the pulsar wind, the shock-cone wraps
 around  the pulsar. Around the INFC ($\sim -150$days),
  the post-shocked pulsar wind moves
  toward the Earth   and hence enhances the
  emissions. These are suppressed at around the SUPC
  ($\sim +10$days) where the post-shocked pulsar wind moves away
  from the Earth. However, the Doppler boosting effect will not be
  the reason for the rapid
  increase in the X-ray emissions observed by the Swift.

 \subsubsection{Dependence  on wind momentum ratio} 
\label{etade}
 \begin{figure}
   \centering
   \epsscale{1.0}
   \plotone{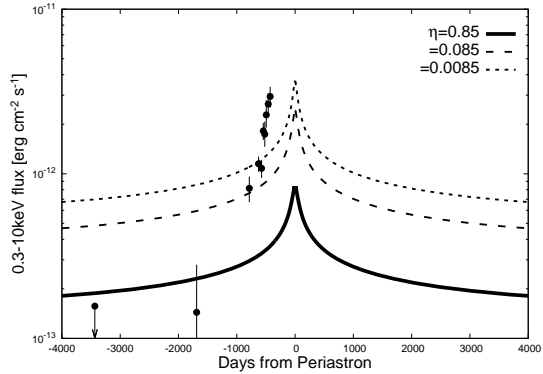}
   \caption{The expected  orbital modulations of the X-ray
     with the different momentum ratio
     of the pulsar wind and stellar wind; $\eta=0.85$  for the solid line,
     0.085 for the dashed line and 0.0085 for the dotted line, respectively.
     The results are for $\sigma_0=0.1$, $\Gamma_{PW,0}=10^4$ and
     $\theta_E=60^{\circ}$. }
   \label{eta1}
 \end{figure}
 \begin{figure}
   \centering
   \epsscale{1.0}
   \plotone{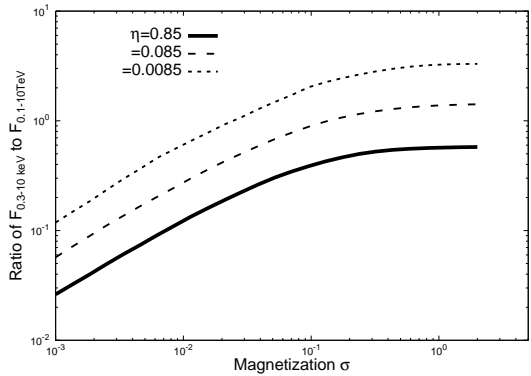}
   \caption{The calculated
     ratio of the X-ray flux and TeV flux at the periastron
     as a function of the magnetization parameter. The results are for
     the momentum ratio of the two winds $\eta=0.0085$ for the solid line,
     $0.085$ for the dashed line and $0.85$ for the dotted line, respectively.
     In addition, $\Gamma_{PW,0}=10^4$ and $\theta_E=60^{\circ}$. }
   \label{depend1}
    \end{figure}

    It has been considered that
   mass loss from the companion star will be driven by the radiation pressure, and
   the rapidly rotating main-sequence star,
   such that the angular velocity of the star rotation
   is close to its  Keplerian angular velocity, and produces a stellar wind enhanced in
    the polar region (Georgy et al. 2011 and reference therein).
    It is also considered that anisotropy of the pulsar wind explains  the torus-like and jet-like
     structures   of the pulsar wind nebulae, and
  theoretical models suggest that  angular  distribution of the pulsar wind energy is
  proportional to $\sin^2\theta$, where $\theta$ is the angle measured from the spin axis (Bogovalov
  and Khangoulian 2002). The anisotropy of the two winds could change the momentum ratio $\eta$ at the
  their interaction region  along the orbit. 

  Besides the large structure of the anisotropy, an irregularity of the stellar wind from the high-mass star could be formed.   
  It has been discussed that a radiatively non-stationary acceleration process produces
  the clumping  (small-scale density inhomogeneity) of the stellar wind from  the high-mass star
  (Runacres \& Owocki 2002: Owocki \& Cohen 2006). It was analyzed that the typical size of a clump at
  the stellar surface is $\sim 0.01R_*$, and
  may linearly expand with the radial distance.  A larger structure of clumps could be formed in the wind due to difference
  production mechanisms, magnetic field inhomogeneities, and star's rotation/pulsation (see Bosch-Ramon 2013).
  Although the clumps fill only a small fraction of the volume
  of  the wind region, it is thought that they carry most of mass ejected from the star, and have a mass density larger than
  the average density of the wind. Studies with hydrodynamic simulations have shown  that
  the clumping could develop  in size and mass density due to merging between dense shells as traveling from
  the star,   although it was also pointed out that a  break-up of the dense shell 
  as a consequence of the Rayleigh-Taylor or thin-shell instability limits the clump growth (Bozzo et al. 2016, and references therein). 
  If a clumpy wind is formed in a high-mass binary system hosting a  compact object,
  it is expected that the irregularity of the mass density in the stellar wind
  will cause a temporal evolution of the momentum ratio ($\eta$), yielding the temporal variations in the shock emissions
  (Owocki et al. 2009; Bosch-Ramon 2013; de la Cita et al. 2016).

 The calculated X-ray fluxes depend
 on the wind momentum ratio, which determines the location of the shock.
 Figure~\ref{eta1} shows the calculated 
   light curves for  $\eta=0.85$ (solid lines), 0.085 (dashed lines) and 0.0085 (dotted lines), 
   respectively, with the parameters
   of  $\sigma_0=0.1$ and $\Gamma_{PW,0}=10^4$.
   As we can see in Figure~\ref{eta1}, the calculated X-ray flux increases
   with decreasing the momentum ratio, because the shock distance from
   the pulsar decreases and hence the  magnetic field at the shock
   increases with  decreasing the momentum  ratio. As the figure shows,
   a  decrease in the momentum ratio (that is, the stellar wind becomes stronger relative to the pulsar wind) by a factor of ten can give an increase in 
   the observed  X-ray fluxes during $\sim -2000$days to $\sim -1000$days from the periastron.
   Such momentum evolution with time could be caused by the anisotropy of the stellar wind/pulsar wind. A shorter time scale variability will be also
   possible as a consequence of the pulsar wind/clumpy wind interaction.
   
The observed flux ratio between the X-ray and TeV  
energy bands  may provide an additional information for the magnetization parameter and momentum  ratio of the pulsar wind to the stellar wind. 
 As we expect,  the flux ratio of the X-ray 
 and TeV does not greatly  depend on $\Gamma_{PW,0}$, but it is sensitive to
 the energy densities of the magnetic field and
 soft-photon field at the emission region,
 which mainly depends on the magnetization parameter and the momentum ratio 
of the two winds. Figure~\ref{depend1} presents  the dependence of 
$F_{0.3-10{\rm keV}}/F_{0.1-10{\rm TeV}}$ on 
the magnetization parameter and the wind momentum ratio.
 If the future observations  could measure the flux ratio 
in X-ray and TeV energy bands, we could   discuss the momentum ratio 
of the two winds; for example,  if the future observations provide
$F_{0.3-10{\rm keV}}/F_{0.1-10{\rm TeV}}\sim 1$ at the periastron,
the current model predicts that the magnetization $\sigma_0\sim $1 if
the momentum ratio  $\eta\sim 0.0085$.

\subsubsection{Radial-dependent $\sigma$ of the pulsar wind}
\label{radial}
\begin{figure*}
\centering
\epsscale{1.0}
\plotone{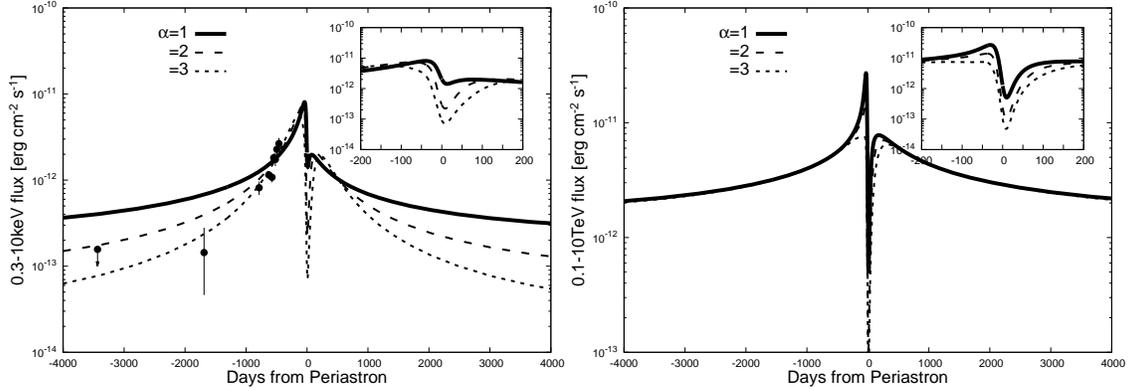}
\caption{The X-ray (left) and TeV (right) light curves with a radial dependent magnetization
  parameter. The solid, dashed and dotted lines are results for $\alpha=1$, 2 and 3, respectively
  in equation~(\ref{sr}). All cases assume  $\eta=0.02$ and  $\Gamma_{PW,0}=10^4$. The magnetization parameter for each case is normalized  so as to be $\sigma\sim 0.01$ at $r\sim 2.5$AU.}
 \label{fit}
 \end{figure*}

\begin{figure}
\centering
\includegraphics[width=0.5\textwidth]{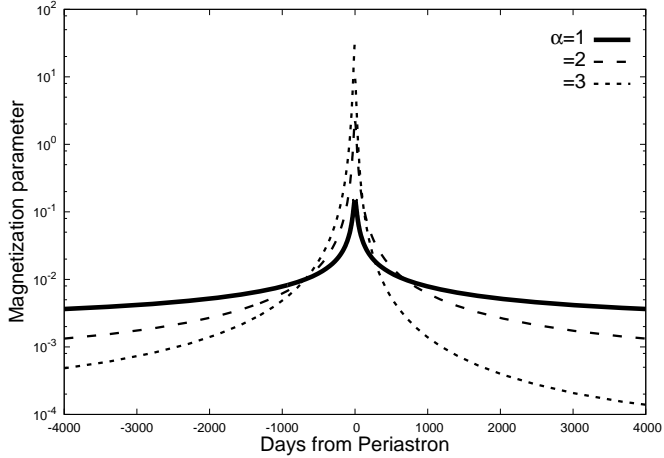} 
\caption{The value of the magnetization parameter at the shock-apex along the orbital phase.
Model parameters for the lines are same as those in  Figure~\ref{fit}.}
 \label{sigma}
 \end{figure}

As seen in Figures~\ref{light}, \ref{light-d} and \ref{eta1}, the calculated
X-ray light curve  cannot explain  the Swift observations
after $\sim -2000$days, if the momentum ratio does not significantly
change along the orbit. The radial-dependent magnetization parameter has
been discussed in the previous studies
to explain the orbital modulations of the  gamma-ray
binaries (Takata \& Taam 2009; Kong et al. 2011; Takata et al. 2014a).
In this paper, we explore the radial dependence with a function form of  
\begin{equation}
  \sigma (r)\propto r^{-\alpha}.
  \label{sr}
\end{equation}

The left panel in Figure~\ref{fit}  shows the results of the fitting for the Swift X-ray data with
different power law indices; $\alpha=1$ for solid line, 2 for dashed line and 3 for dotted line.
Other parameters are $\eta=0.02~(\dot{M}\sim 4\times 10^{-8}M_{\odot}$/yr) and  $\Gamma_{PW,0}=10^4$.
In addition, we applied $v_{pw}=v_{pw,2}$ taking into account the Doppler boosting effect discussed in
section~\ref{dop}.  To explain the observed flux level from   Swift,
we normalized the magnetization parameter so as to be $\sigma\sim 0.01$ at $r\sim 2.5$aU for each  power law index. Figure~\ref{sigma} shows the
evolution of the fitting magnetization parameters
at the apex of the shock cone as a function of the orbital phase.

We can see in the Figure~\ref{fit} that a faster evolution (larger $\alpha$)
of the magnetization parameter with the radial distance reproduces
a X-ray light curve being  more consistent with the whole Swift
observations after $-4000$days. As the figure shows,
the current model predicts that the peaks of the flux in the light curve
occurs  at a day before the periastron and the flux then rapidly
decreases during   $\sim -20$days  and $\sim +10$days by about
one order of magnitude.  This
is because (1) the assumed magnetization parameter determined by
the relation, $\sigma(r)\propto r^{-\alpha}$, exceeds  unity for $\alpha>2$,
as Figure~\ref{sigma} shows, and (2) the Doppler boosting suppresses 
the emissions around the SUPC. This feature
can be tested  by the future observations.

\subsection{GeV emissions and multi-wavelength spectra}
\begin{figure}
\centering
\includegraphics[width=0.5\textwidth]{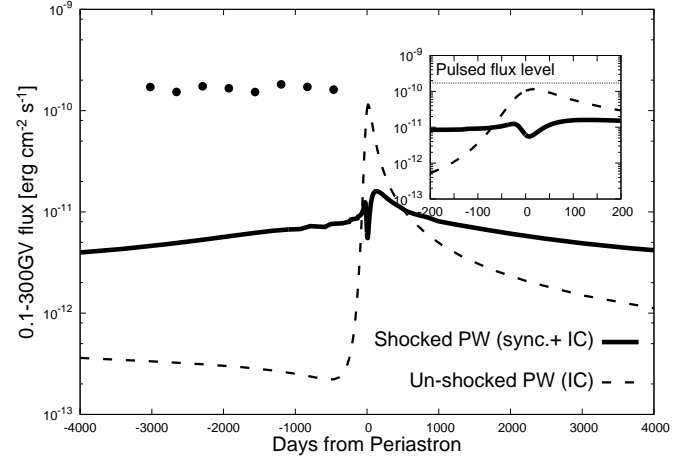} 
\caption{The orbital modulation of the calculated 0.1-100GeV fluxes. 
The solid line and dashed line correspond to the emissions from 
the shocked pulsar wind and cold-relativistic pulsar (un-shocked pulsar wind), 
respectively. The shock emissions include  the synchrotron radiation and ICS 
process of  the shocked pulsar wind emissions.
The 8-years Fermi-LAT data are also displayed. 
The results of calculation are for $\Gamma_{PW,0}=10^4$,
 $\sigma_0=0.1$ and $\theta_E=60^{\circ}$.}
\label{gevlight}
 \end{figure}
Gamma-ray binaries have also been detected by the Fermi-LAT and 
the emission will originate  from 
the  magnetospheric processes and/or the intra-binary processes.  
For PSR~J2032+4127, the pulsed GeV emission due to the pulsar spin 
 has been measured by the Fermi-LAT with an energy 
 flux $\sim 1.6\times 10^{-10}{\rm erg~cm^{-2}~s^{-1}}$
 (section~\ref{observations}).  The high-energy
 tail of the  synchrotron spectrum and the low energy tail of  the  ICS process
of the intra-binary shock  discussed in the previous sections could 
contribute to the energy bands  (0.1-300GeV) of the Fermi-LAT. In addition 
to the shock emissions, the ICS process of the cold-relativistic 
pulsar wind with $\Gamma_{PW,0}\sim 10^{4-5}$ produces GeV gamma-rays (Figure~\ref{cpw}).

Figure~\ref{gevlight} compares  the observed flux level of 
the pulsed emissions with  the predicted flux levels of the shocked 
emissions (solid line) and  of the ICS process of the cold-relativistic 
pulsar wind (dashed line). In the calculation, we assume $\Gamma_0=10^4$ for 
the average Lorentz factor of the particles of the cold-relativistic pulsar
wind, and $\theta_E=60^{\circ}$ for the inclination of the system.
As we see in Figure~\ref{gevlight}, the emissions from the
intra-binary space are  comparable to the magnetospheric emissions around
the periastron.
With the current  parameters, therefore, we  expect that  the Fermi-LAT 
measures the orbital modulation of the GeV flux during the
periastron passage that will occur in late 2017 or in early 2018.

\begin{figure*}
\centering
\epsscale{1.0}
\plotone{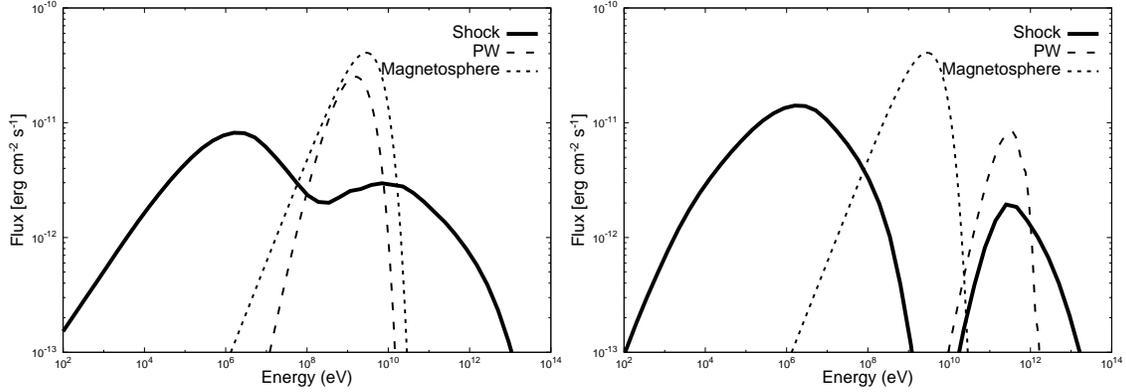}
\caption{The multi-wavelength spectra averaged over during $-100$days
  and $+100$days from
  the periastron and calculated with $\Gamma_{PW,0}=10^4$ in the left 
  panel and $\Gamma_{PW,0}=10^6$ in the right panel, respectively. 
  The calculations assume the momentum
  ratio   $\eta=0.02$,
  the power law index $\alpha=2$ in equation~(\ref{sr}), the post-shock velocity  $v_{pw}=v_{pw,2}$ and
  the Earth viewing angle $\theta_E=60^{\circ}$.   The solid, dashed and 
dotted line correspond to the emissions from the shocked pulsar wind, 
from the cold-relativistic pulsar wind and the pulsed emissions, respectively. 
For the spectrum of the pulsed emissions, we apply the observational result
in the  Fermi-LAT pulsar catalog (2013).}
\label{spectrum}
 \end{figure*}

In Figure~\ref{spectrum}, we present the model  broadband spectra
averaged over $-100$days to $+100$days; the model parameters are
$\alpha=2$ in equation~(\ref{sr}), $\eta=0.02$ and
$v_{pw}=v_{pw,2}$, which reproduce
the observed X-ray light curves. To see the dependence of the
Lorentz factor $\Gamma_{PW,0}$, we calculated the spectra for
$\Gamma_{PW}=10^4$ (right panel) and $10^{6}$ (left panel), respectively.
The solid line shows the calculated
spectrum combining  the synchrotron and ICS 
process at the shock, and the dashed line represents the spectrum 
of the ICS process of the cold-relativistic pulsar wind.
The current model predicts that
the magnetospheric emissions  dominate
in the spectral energy distribution averaged
over the periastron passage. If the minimum Lorentz factor
of the shocked pulsar wind particles is $\Gamma_{PW}\sim 10^6$,
the TeV spectrum will show a turnover at around
$\sim 10^{11} {\rm eV}$ and the X-ray spectrum  has a break at  around 10keV.

\section{Discussion and summary}
\label{discussion}
\subsection{Emissions from the secondary pairs}
\begin{figure}
  \centering
  \epsscale{1.0}
  \plotone{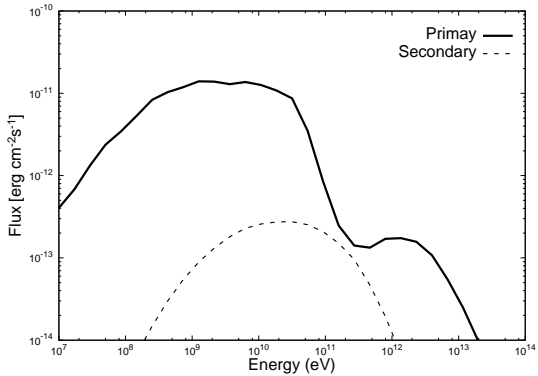}
  \caption{A comparison between the spectra of the primary (solid line) and secondary (dashed) ICS process observed on the Earth.
  The results are for $\sigma_0=0.1$, $\Gamma_{PW,0}=10^4$ and at the periastron. }
  \label{cascade}
\end{figure}
During the periastron passage,   the TeV photons from the shocked pulsar wind may be converted into
pairs by the photon-photon pair creation process with the stellar photon
fields; in particular, the TeV photons emitted toward the B star will be totally absorbed.
These secondary pairs created in the stellar wind side
will also emit the non-thermal photons via the synchrotron  and ICS processes, and may
initiate a further pair-creation cascade.  If these pairs are isotropized,
they will emit the photons propagating  toward the observer, even though the primary TeV photons
do not do this (Bednarek 1997, 2007; Sierpowska \& Bednarek 2005, 2008; Sierpowska-Bartosik \& Torres 2007, 2008;
Yamaguchi \& Takahara 2010; Cerutti et al. 2010).

The emissions from the secondary pairs produced in the stellar wind side
will be dominated by the
 ICS process. The ratio of the radiation power between the synchrotron radiation and ICS processes will be described by
 the energy density ratio of the magnetic field to the stellar
photon. The ICS process dominates the synchrotron radiation process at  the radial distance from the star of  
\begin{equation}
  R\ge R_{*}\times 6.5^{\frac{1}{2(m-1)}}
    \left(\frac{B_{*}}{10^{3}{\rm G}}\right)^{\frac{1}{m-1}}
    \left(\frac{T_{*}}{30000{\rm T}}\right)^{-\frac{2}{m-1}},
\end{equation}
where $B_*$ represents the stellar magnetic field at the stellar surface. In addition, we assumed  $B(R)\propto R^{-m}$ as
the radial evolution of the stellar magnetic field. It is known
that some high-mass main-sequence stars  have a surface  dipole magnetic field of order of $10^3$G (Walder et al. 2012).
With a typical value of $m=2\sim 3$, we can see that
the synchrotron radiation of the created pairs in the stellar wind side
can be important only at very near the stellar surface and its contribution
to  th observed X-ray emissions will be negligible. 

By tracing the primary TeV gamma-rays from all emission regions considered
in the calculation box, we calculated  the ICS process of the secondary pairs.
We assumed that  the pairs produced in the stellar wind side are quickly
isotropized, and we  calculated the emission process with a constant
photon field during the crossing time,  $\sim R_i/c$,
where $R_i$ represents the distance of the pair-creation position
from the stellar surface.
Figure~\ref{cascade} compares the spectra, which are  measured on  Earth
of the ICS photons from the shock region (solid line)  and from the secondary pairs (dashed line)
at the periastron, where we assumed that all photons from the secondary pairs
can escape from the pair-creation
process. A  further pair-creation cascade process will make the  spectra softer.
As the figure shows,  the integrated flux from the
secondary pairs is much less than the primary emissions, and therefore we expect that
the contribution of the emissions from higher-order pairs is  much less than
that from the shock emissions. The emissions from the secondary pairs could contribute to  the 0.1-1TeV bands,
where the shock emissions around the periastron are significantly absorbed.

\subsection{Pulsar/Be disk interaction}
\begin{figure*}
  \centering
  \epsscale{1.0}
  \plotone{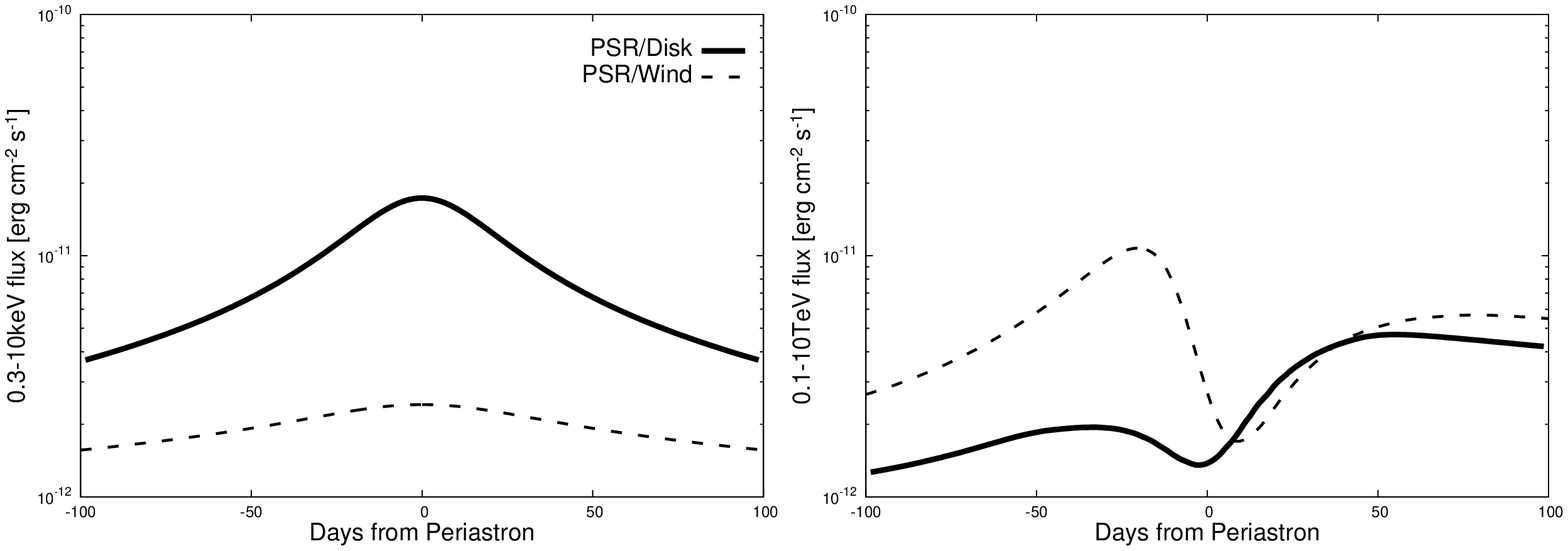}
\caption{The solid lines show emissions from the pulsar
 wind stopped by the Be disk at the periastron passage. 
The calculations assume that  
all pulsar wind ($4\pi$ in the solid angle) is stopped by the radial 
distance from the pulsar, which is  given by equation~(\ref{shockdisk}).
For comparison, the dashed lines show the calculated light curves of the emissions
from the shock due to the pulsar wind/stellar wind 
interaction with $\eta=0.085$.  }
\label{diske}
 \end{figure*}
The interaction of PSR~J203+4127 and the Be disk of MT91~213 
may enhance the high-energy emissions,  as in the case with
the PSR B1259-69/LS2883 system showing
an  increase in the X-ray/TeV fluxes at the pulsar/Be disk interaction 
phase (Chernyakova et al. 2006).  As we discussed in section~\ref{disk}, we expect that 
the pulsar/Be disk interaction 
in this system affects the observed emissions, provided that  the base 
density is $\rho_{0}>10^{-10}{\rm g~cm^{-3}}$  and
the interaction occurs  at the periastron passage, say between 
$-100$days and $+100$days from the periastron (Figure~\ref{bow}).
Since we do not know the geometry (e.g. 
inclination relative to the orbital  plane) of the disk, we cannot 
predict when the pulsar interacts with it. We expect however 
that the pulsar/Be disk interaction will occur  at 
least once during the periastron passage during $-100$days and $+100$days, 
since the true anomaly  $\pm90^{\circ}$ 
 corresponds to  $\sim \pm 50$days from the periastron.

If the base density of the Be disk is close to 
$\rho_0\sim 10^{-9}{\rm g~cm^{-3}}$, an interaction between 
pulsar and Be disk at the periastron passage will form a  cavity in  
the pulsar wind, and most of the pulsar wind ($4\pi$ sr in solid angle) 
will be stopped  at the radial distance from the pulsar 
given by equation~(\ref{shockdisk}) and Figure~\ref{bow}. 
In Figure~\ref{diske}, we show an  example of the calculated light curve
for the pulsar/Be disk interactions  and compare the estimated
X-ray/TeV fluxes with  those for the pulsar/Be wind interactions. For 
the calculation, we assumed the base density $\rho_{0}=10^{-9}{\rm g~cm^{-3}}$,
the magnetization $\sigma_0=0.1$ and the momentum  ratio $\eta=0.085$
for the pulsar wind/stellar wind interaction.

Figure~\ref{diske} indicates that the pulsar/Be disk interaction predicts
 X-ray fluxes several factors larger  than those  due to  
 the pulsar/Be wind interaction, if the magnetization parameter does not
 evolve with the radial distance.
 Within the framework of the current calculation, the shock distance from the 
pulsar is closer,  and the magnetic field at the shock is stronger 
for the pulsar/Be disk interaction, in which $r_s/a\sim 0.025$ at periastron, 
than the pulsar/Be wind interaction, 
for which $r_s/a\ge \eta^{1/2}/(1+\eta^{1/2})\sim 0.22$ with $\eta=0.085$. 
As a result, the synchrotron power is stronger for the pulsar/Be disk 
interaction than the pulsar/Be wind interaction.  This result suggests 
that if the base density of the Be disk is $\rho_0\sim 10^{-9}{\rm g~cm^{-3}}$, 
the interaction of pulsar/Be disk may give rise to local maximum in the orbital 
modulation of the X-ray emissions. 

As the right panel in Figure~\ref{diske} shows, the calculated TeV flux 
for the pulsar/Be disk interaction is lower than that for the pulsar/Be wind
interaction at around the periastron.
 This is because the synchrotron cooling time scale for TeV 
leptons becomes  shorter than the  ICS cooling and adiabatic cooling time scales
 at the shock for the pulsar/Be disk interaction.
 At the periastron, for example, the shock distance at the apex 
for the pulsar/Be wind interaction is $r_s\sim 0.22a$, for which the 
inverse-Compton cooling and adiabatic cooling time scales of TeV electrons 
are  shorter than the synchrotron cooling time scale (Figure~\ref{cool}).
 For pulsar/Be disk interaction, the shock distance  becomes 
 $r_s\sim 0.0025a$,  which increases the magnetic field at the shock  by
 a factor of $\sim 10$,  and hence decreases the synchrotron cooling time scale
 by about  $\sim  1/100$.  As a result, 
 the ICS cooling  scale, which  is less sensitive to the shock distance, 
 becomes longer  than the synchrotron time scale, and  the ICS emissivity
 is suppressed. 

\subsection{Formation of a  disk around the pulsar}
As we discussed in section~\ref{disk}, if the base density of the disk 
is high enough, the pulsar wind will be confined within a small region 
by the Be disk matter. When the radius of the cavity estimated in 
Figure~\ref{bow} is less  than the radius below which the kinetic energy 
of the disk gas is less than the gravitational potential energy of 
the pulsar, the disk matter may  be gravitationally captured by the 
pulsar, and results in the formation of a  disk around the pulsar. 
The capture radius measured from the pulsar may be  estimated as 
\begin{equation}
r_{cap}\sim \frac{2GM_{N}}{v_{r}^2}
=0.25{\rm AU}\left(\frac{M_N}{1.4M_{\odot}}\right)
\left(\frac{v_{r}}{10^{7}{\rm cm~s^{-1}}}\right)^{-2},
\end{equation}
where $M_{N}$ is the mass of the pulsar and $v_r$ is the relative 
velocity of the pulsar with respect to the disk matter. Near
the periastron, $a\sim 1$AU, the orbital  velocity of the pulsar is
$v_{p}\sim 10^7{\rm cm~s^{-1}}$. The velocities of the Kepler motion and
the radial velocity of the Be disk at the pulsar position are of the order of 
$v_{d,K}(1{\rm AU})\sim v_{d,K}(R_*)(R_*/1{\rm AU})^{1/2}\sim 10^7{\rm cm~s^{-1}}$ and $v_{d,r}\sim 0.1c_s\sim 10^5{\rm cm~s^{-1}}$, respectively
(Okazaki et al. 2011). Hence, we expect that
the relative velocity is of order of $v_r\sim 10^{7}{\rm cm~s^{-1}}$,
although this  depends on the disk geometry and the rotation direction.

As Figure~\ref{bow} shows, 
 at near the periastron, where the separation is $a\sim 1$au, the estimated 
shock distance due to the pulsar/Be disk interaction is of order 
of $r_s\sim 0.1$ au for $\rho_{0}\sim 10^{-9}{\rm g~cm^{-3}}$, suggesting  
the disk matter could  be captured by the pulsar because $r_s<r_{cap}$.
The accretion rate  may be estimated as 
\begin{eqnarray}
{\dot M}_{acc}&\sim& f\rho(a)r_{cap}^2v_{r}
\sim 4\times 10^{17}{\rm g~s^{-1}} \nonumber \\
&\times &f\left(\frac{\rho(a)}
{3\cdot 10^{-15}{\rm g~cm^{-3}}}\right)\left(\frac{r_{cap}}{0.25{\rm AU}}\right)^2\left(\frac{v_r}{10^7{\rm cm~s^{-1}}}\right)
\end{eqnarray}
where $f\sim 1 $ is the geometrical factor and we used 
the parameter at $a\sim 1.5$AU (typical separation during the periastron
passage) with  $\rho_0=5\times 10^{-10}{\rm g~cm^{-3}}$.  We note that 
for the stellar wind, $v_r\sim 10^{8}{\rm cm~s^{-1}}$, the capture radius 
becomes $r_{cap}\sim 0.0025$AU, which is usually smaller  than 
the shock radius due to the pulsar wind and stellar wind interaction
(Figure~\ref{shock}).  Hence the stellar wind will not be captured by
the pulsar.

There will be an relative angular 
momentum of the captured matter with  respect to the pulsar, indicating 
the captured matter forms a circular orbit around the pulsar at a
distance ($\equiv r_{circ}$)
where the relative angular momentum per unit mass 
is equal to the angular momentum
of the Kepler orbit around the pulsar, $(GM_Nr_{icrc})^{1/2}$.
The angular frequency of the orbital motion of the pulsar is
$\omega_N\sim v_{p}/a \sim 5\times 10^{-7}{\rm rad~s^{-1}}$.
The angular frequency of the circular  orbit of the Be disk around
the B-star is also  $\omega_{d}\sim \omega_{N}$, because $v_{d,K}\sim v_{p}$.
Roughly speaking, therefore, the angular frequency associated with  the
relative angular momentum will be of the order
of $\omega\sim 5\times 10^{-7}{\rm rad~s^{-1}}$, although again
it depends on the disk geometry and rotation direction.
The circularization  radius of the captured matter 
may be  estimated as  (Frank et al. 2002) 
\begin{eqnarray}
  r_{circ}&\sim &\frac{r_{cap}^4\omega^2}{16GM_{N}} 
  \sim1.7\times 10^{10}{\rm cm} \nonumber \\
 &\times &  \left(\frac{M_N}{1.4M_{\odot}}\right)^3
\left(\frac{\omega}{5\cdot10^{-7}{\rm rad~s^{-1}}}\right)^2
\left(\frac{v_{r}}{10^7{\rm cm~s^{-1}}}\right)^{-8}.
\label{circ}
\end{eqnarray}

If we apply  the standard Shakura-Sunyaev disk model, the disk matter 
will move inward with a dynamical time scale 
 $\tau_d(r)\sim r/v_{d,r} \sim 15{\rm days}\alpha^{-4/5}_{0.1}{\dot M}^{-3/10}_{17}
r_{10}^{3/4}$, where $\alpha_{0.1}$ is the viscous parameter in units of 0.1,
suggesting the accreting  matter will take several weeks
to reach  the pulsar after the capture event. If the dynamical
time scale ($\tau_{d}$)
is longer than the pulsar's disk crossing time scale,
which may be of the order of $\tau_c\sim 2H/v_p\sim 4{\rm days}
(H/0.1{\rm AU})(v_p/10^7{\rm cm~s^{-1}})$, the accretion disk may not develop
around the pulsar.  We note however that the circularization 
radius given by equation~(\ref{circ}) is sensitive to the relative velocity;
 with $v_{r}\sim 2\times 10^7$cm, for example,
 $r_{circ}\sim 6\times 10^{7}$~cm, which is inside of 
the light cylinder. Moreover, we do not know the geometry of the disk. If 
the disk  plane is parallel to the orbital plane, the pulsar/Be disk 
continues to interact during the periastron passage. Hence it is possible 
that the accretion disk develops with a radial length $\sim 10^{10}{\rm cm}$ 
  around the pulsar at the periastron passage. 

\begin{figure}
  \centering
  \epsscale{1.0}
  \plotone{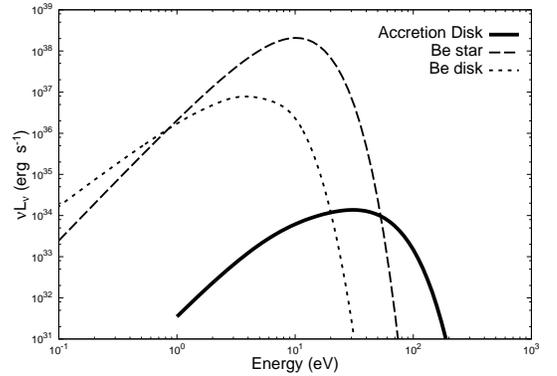}
  \caption{IR/optical/UV emissions from  PSR~J2032+4127/MT91~213 system.
    Solid line: Emissions from the Shakura-Sunyaev disk with $\dot{M}=4\times 10^{17}{\rm g~s^{-1}}$ 
and the disk extends from $r=7\times 10^8$cm to $10^{10}$cm. Dashed line: 
Black body emission from the stellar surface with the temperature
 $T_{*}=30000$K. We assume the Planck function. Dotted line: Black body 
radiation from the isothermal Be disk with the temperature  $T_d=0.6T_{*}$ 
(Carciofi \& Bjorkman  2006).
}
\label{bb}
\end{figure}

  If the accretion disk forms around the pulsar, the disk supplies
  UV photons  in the pulsar wind region.
  Figure~\ref{bb} summarizes  the radiation power ($\nu L_{\nu}$)
  for  the accretion disk (solid line), stellar  surface (dashed line), 
  and Be disk (dotted line).  For the accretion disk,
  we assume an  accretion rate $\dot{M}=4\times 10^{17}{\rm g~s^{-1}}$
 and the disk extends from the light cylinder $r=r_{lc}$ of the pulsar to $r=10^{10}{\rm cm}$.
For the stellar emission, we assume the Planck function with $T_*=30000$K 
and $R_*=10R_{\odot}$. For the Be disk emissions, we assumed  
the disk temperature $T_d=0.6T_*$ after  the calculation of  
 Carciofi \& Bjorkman (2006). We ignore the effects of the emission 
and absorption lines on the spectra.

\begin{figure*}
\centering
\epsscale{1.0}
\plotone{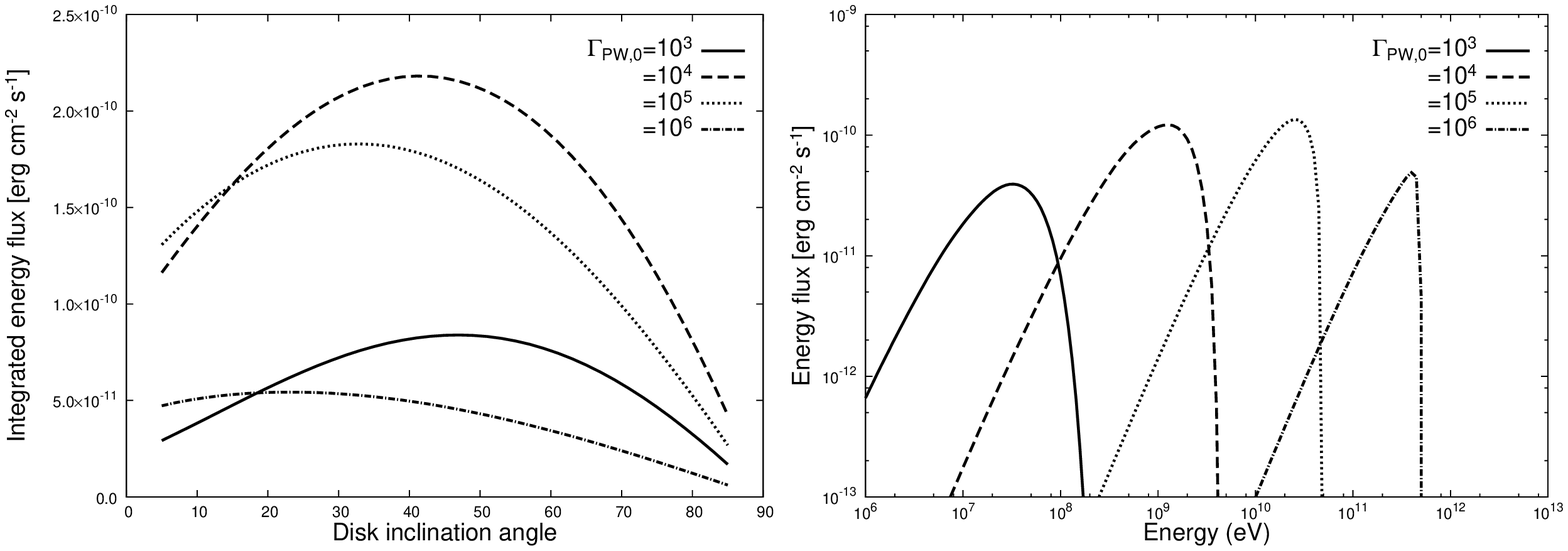}
\caption{The calculated ICS
  process between the cold-relativist pulsar wind 
and the soft-photons from the accretion disk around the pulsar. 
Left: The integrated 
flux for the different Lorentz factor of the pulsar wind as a function 
of the disk inclination angle, which is the angle 
 between the Earth direction and direction perpendicular to the disk plane. 
In the calculation, the disk extends from $r=r_{LC}$ to $r=10^{10}$cm. 
Right: the calculated spectra for the disk inclination angle 
$\theta_i=45^{\circ}$.}
\label{cpdisk}
\end{figure*}

As discussed in above, the captured matter 
 will take about several weeks to reach the pulsar magnetosphere.  
Before the accretion disk reaches to the light cylinder radius, the rotation 
activity of the pulsar is still turned on and produces the cold-relativist 
pulsar wind.  Then we can expect that the ICS process of the
cold-relativistic pulsar will boost up the UV photons  from the disk to the  higher energy photon.
Figure~\ref{cpdisk} summarizes  the calculated spectra (left panel)
and the integrated flux as a function of the accretion disk inclination angle (right panel).
The figure shows the results when the accretion disk extends from 
the light cylinder radius $r=r_{lc}$ to $r=10^{10}{\rm cm}$.  In addition, we assumed 
the mono-energetic distribution of the particles 
in the cold-relativistic pulsar wind. We took into account the 
energy loss of the pulsar wind due to the ICS process. As Figure~\ref{cpdisk} 
shows,  if the initial 
Lorentz factor is $\Gamma_{0}=10^{4-5}$ which  produces
the scattered photons with 1-10GeV energy,  the calculated  flux level can 
reach an  order of $2\times 10^{-10}{\rm erg~cm^{-2}~s^{-1}}$, which is 
comparable to that of the pulsed emissions.

  Pulsar binary systems may provide a valuable laboratory to study the accretion process on
  rapidly spinning and strongly magnetized neutron stars. One may consider that
  once the accretion disk crosses the light cylinder, the rotation-powered activity  would be quenched.
  This  is because the  copious plasma fills the acceleration region at  the rotation-powered 
stage, where the charge density  deviates  
from the so-called Goldreich-Julian charge density (Goldreich \& Julian 1969).
However, the recent observation suggests the rotation-powered activities and the accretion disk can co-exist in the magnetosphere.  
A transient-millisecond pulsar is a  binary millisecond pulsar with
low mass companion star, and it transits  between
the rotation-powered stage and the accretion stage (Archibald et al. 2009).
PSR~J1023+0038 transited from  its rotation powered
stage to the accretion  stage at the latter in  2013 June, when 
the pulsed radio emission from this pulsar disappeared (Stappers et al. 2013).
Coherent pulsed X-ray emission discovered  after 2013 June suggests
an accretion of matter on the neutron star surface (Archibald et al. 2015).
Jaodand et al. (2016), however, found  that  the spin-down rate in the accretion stage is  higher than  that  in the rotation-powered stage,
and they concluded that
the rotation activity is still operating   in the accretion stage.  Enhancement of observed  GeV emission
in  the accretion stage also suggests the survival of the rotation activity  in the
accretion stage
(Takata et al. 2014b).  This model was also applied to  another transient-millisecond pulsar PSR J1227-4853 (Bednarek 2015).
Gamma-ray binaries with a Be companion may also offer
opportunities to study this  accretion process on young pulsars.

The ICS model of the pulsar wind  could provide an explanation for
the origin of the flare-like 
GeV emissions from  PSR B1259-63/LS 2883 system after the second Be disk 
 passage of the pulsar. It has been observed that the GeV peak in the orbital 
modulation occurs  $\sim 20$ days after the X-ray peak 
(Tam et al. 2011, 2015). It has been considered that the phase of the X-ray peak
 corresponds to that  of the pulsar/Be disk interaction. So we may expect 
 that the Be disk matter starts to be captured by the pulsar
 around the orbital phase of the X-ray peak. 
 Since it will take several weeks to fully develop the accretion disc around
 the pulsar, the GeV  peak position  will delay from   the X-ray peak position.

Finally, the magnetic radius $r_{M}$ from the pulsar is defined as the
 distance where the dipole magnetic field of the pulsar
 begins to dominate the 
dynamics of the accreting matter, and it is estimated from 
$B^2(r_{M})/8\pi=\rho_{acc}v^2$ with $\rho_{acc}v=\dot{M}/4\pi r_{M}^2$ and 
$v\sim (2GM_N/r_M)^{1/2}$,
\begin{eqnarray}
  r_{M}&\sim& 3\times 10^8{\rm cm}\left(\frac{B_s}{10^{12}{\rm G}}\right)^{4/7}\nonumber \\
  &\times &
\left(\frac{\dot{M}}{10^{17}{\rm g~cm^{-3}}}\right)^{-2/7}
\left(\frac{M_N}{1.4M_{\odot}}\right)^{-1/7}.
\end{eqnarray}
This radius is smaller than the light cylinder radius of PSR~J2032+4127 
($r_{lc}=6.8\times 10^8$cm).  The co-rotation radius ($r_{co}$) 
is defined as the distance where the angular velocity of  
Keplerian motion is equal to the spin angular velocity of the pulsar,
 namely 
\begin{equation}
r_{co}=\left(\frac{GMP^2}{4\pi^2}\right)^{1/3}
\sim 4.6\times 10^{7}{\rm cm}\left(\frac{P}{0.143{\rm ms}}\right)^{2/3}
\left(\frac{M_N}{1.4M_{\odot}}\right)^{1/3},
\end{equation}
which will be smaller than the magnetic radius, suggesting
the accretion process is in propeller regime,
and not all infalling  matter will accrete
on the pulsar surface.  The outflow due to  the propeller effect also 
could cause  the high-energy emission
 (e.g. Papitto \& Torres, 2015).

In summary, we analyzed the Swift and Fermi-LAT data of the PSR~J2032+4127/MT91~213
binary system, which is a   candidate to be  another TeV  gamma-ray binary.
The X-ray flux has rapidly increased as the pulsar approaches
periastron, which will occur  in late 2017 or early 2018, while
the GeV flux measured by Fermi-LAT shows no significant
change. We investigated  the X-ray emission by
the synchrotron process of the pulsar wind particles accelerated at
 the shock due to  the pulsar wind and stellar wind interaction. 
 We argued that the increase in the observed X-rays is caused by
 (1) variation of the momentum ratio of the two winds along the orbital phase or  (2) the  evolution
 of  the magnetization parameter of the pulsar wind
 with the radial distance as  $\sigma(r)\propto r^{-\alpha}$
 and  $\alpha\sim 2-3$. The current model predicts
that the peak fluxes in the 0.3-10keV and 0.1-10TeV energy bands are
of the order of $\sim 10^{-11}{\rm erg~cm^{-2}s^{-1}}$.
The gamma-ray binary system could  provide a unique laboratory to study  the emissions from
  the cold-relativistic pulsar wind. The current model suggests that
  the ICS  process
  of the cold-relativistic pulsar wind could contribute to
  the Fermi-LAT observations at the periastron passage (Figure~\ref{gevlight}).
  For this binary system,
  the pulsar/Be disk interaction will affect to the observed emissions,
  provided that the base density of the Be disk
  is $\rho_0>10^{-10}{\rm g~cm^{-3}}$ and the interaction occurs 
  during  $\sim -100$days and $\sim +100$days. For a Be disk
  with higher base density $\rho_0\sim 10^{-9}{\rm g~cm^{-3}}$,
  the pulsar/Be disk interaction could form an  accretion disk
  around the pulsar. The ICS process of the cold-relativistic
  pulsar wind off the UV from the disk also provides the high-energy
  emission from the pulsar/Be star binary system.
  Future multi-wavelength observations
  will provide a unique opportunity
  to probe the pulsar wind properties and the various emission processes
  caused by the pulsar/Be wind and pulsar/Be disk interactions.

  We express our appreciation to an anonymous referee for useful comments and
  suggestions. We thank A. Okazaki for the useful discussions on the Be disk model. JT is
  supported by NSFC grants of Chinese Government under 11573010 and U1631103.
  CWN and KSC  are supported by GRF grant under 17302315. CYH is supported by the National Research Foundation of Korea through grants 2014R1A1A2058590 and 2016R1A5A1013277. All calculations were done under  the High Performance Computing  Cluster (Hyperion) of the Institute of Particle Physics
   and Astrophysics, HUST.

\end{document}